\documentclass[a4paper,12pt,nohyper]{JHEP3}
\usepackage{amsmath,latexsym,amssymb}
\usepackage{graphicx}
\usepackage[latin1]{inputenc}
\usepackage{subfigure}
\usepackage{nicefrac}
\usepackage{braket}
\RequirePackage{graphicx}

\usepackage{framed}
\def\crbig{\\\noalign{\vspace{1mm}}}

\def\r{\rho}
\def\a{\alpha}

\def\d{\delta}

\def\e{\epsilon}

\def\p{\pi}

\def\th{\theta}

\def\om{\omega}

\def\l{\lambda}

\def\s{\sigma}

\def\cN{{\cal N}}

\def\no{\noindent \vspace{-0.35 cm}\newline}

\def\qq{\qquad}

\def\IR{\relax{\rm I\kern-.18em R}}

\def \ha {{1\over 2}}

\def \ov {\over}

\def\IR{\relax{\rm I\kern-.18em R}}
\def\inv{^{\raise.15ex\hbox{${\scriptscriptstyle -}$}\kern-.05em 1}}

\def\be{\begin{equation}}
\def\ee{\end{equation}}
\def\ba{\begin{eqnarray}}
\def\ea{\end{eqnarray}}

\newcommand{\eqn}[1]{(\ref{#1})}

\title{\boldmath\Large
Holographic approach to deformations \\ of  NS5-brane
distributions  and exact CFTs \unboldmath}

\author{A.~Fotopoulos${}^{1}$,
P.M.~Petropoulos${}^{2}$, N.~Prezas${}^{3}$ and K.~Sfetsos${}^{3,4}$
\\

  \begin{itemize}
  
\item Dipartimento di Fisica Teorica dell'Universit\`a di Torino
and INFN,\\Sezione di Torino,
via P. Giuria 1, 10125 Torino, Italy\\
Email: \email{foto@to.infn.it}
    
  \item     Centre de Physique Th{\'e}orique, Ecole Polytechnique, CNRS--UMR 7644,\\
91128 Palaiseau Cedex, France\\
Email: \email{marios@cpht.polytechnique.fr}
       
\item Theory Unit, Physics Department, CERN,\\
1211 Geneva 23, Switzerland
 \\
Email: \email{nikolaos.prezas@cern.ch}
\item Department of Engineering Sciences, University of Patras,\\
26110 Patras, Greece\\
Email: \email{sfetsos@upatras.gr}
  \end{itemize}

\bigskip

}

\abstract{We consider general planar deformations of a circular
distribution of NS5-branes. The near-horizon region of the latter
admits, after a T-duality transformation, an exact
conformal-field-theory description in terms of the coset model
$SU(2)/U(1) \times SL(2,\mathbb{R})/U(1)$. We derive the exactly
marginal operators corresponding to an infinitesimal planar
deformation using the conjectured holography between the coset
model and the little string theory that resides on the worldvolume
of the NS5-branes. Subsequently, we perform a complementary
analysis of the same deformations using the associated ${\cal
N}=1$ supersymmetric $\sigma$ model and verify the holographic
correspondence. We explicitly demonstrate a precise match between
the two approaches which rests upon a delicate interplay between
exact conformal-field-theory operators and their semiclassical
realizations in terms of target-space variables.}

\preprint{DFTT 26/1007\\
CPHT-RR089.0807\\
CERN-PH-TH 2007/249}

\begin{document}

\setcounter{footnote}{0}
\renewcommand{\thefootnote}{\arabic{footnote}}
\setcounter{section}{0}

\section{Introduction}

Linear-dilaton backgrounds were recognized long ago as string
vacua with rich properties and diverse applications
\cite{Myers:1987fv,Antoniadis:1988vi}. An important step was taken
in \cite{Callan:1991dj}, where the connection between the
linear-dilaton  exact worldsheet theory and the solitonic
target-space objects known as NS5-branes \cite{duff} was
established. Distributions of NS5-branes in their transverse space
generate exact string backgrounds with half supersymmetry broken.
Situations where the underlying $\s$ model can be identified with
a known exact conformal field theory (CFT) are especially
desirable but nevertheless rare. When $k$ parallel NS5-branes are
located at the same point, their transverse near-horizon geometry
is the target space of the $\mathbb{R}_\phi \times SU(2)_{k}$
supersymmetric Wess--Zumino--Witten (WZW) model --
$\mathbb{R}_\phi$ denotes the radial direction that supports the
linear dilaton \cite{Callan:1991dj}. This background exhibits the
${\cal N}=4$ superconformal algebra and its string spectrum was
analyzed in \cite{Kounnas:1990ud}. The only other known case
resulting into an exact CFT description is when the NS5 branes are
uniformly distributed on the circumference of a circle
\cite{Sfetsos:1998xd}. This is also a remarkable theory since the
geometry, two-form and dilaton backgrounds turn out
\cite{Sfetsos:1998xd} to be T-dual to those of the ${SU(2)}/{U(1)}
\times {SL(2,\mathbb{R})}/{U(1)}$ product of conformal cosets.

\no
Over the recent years, NS5-brane distributions attracted further
attention in the advent of holography. In this framework, it is
conjectured that string theory on linear-dilaton-like vacua is the
holographic dual of little string theories (LSTs)
\cite{Aharony:1998ub}. The latter are non-gravitational theories that
capture the dynamics of the worldvolume modes of NS5-branes in
some appropriate decoupling limit.

\no An important consequence of the above holographic duality is
the correspondence between vertex operators of the string theory
on the asymptotic linear-dilaton background and deformations of
the dual string theory \cite{Aharony:2003vk,Aharony:2004xn}.
Hence, starting from a specific NS5-brane distribution, one can
trigger perturbations by giving vacuum expectation values (VEVs)
to appropriate scalar fields defined on their worldvolumes. Such
perturbations amount to displacements of the
 NS5-branes in their four-dimensional
transverse space, and their effect on the underlying two-dimensional worldsheet
 theory can be immediately uncovered using the
holographic dictionary.

\no The present work aims primarily at \emph{demonstrating} the
validity of the above holographic dictionary in situations where
the effect of the perturbations on the locus of the NS5-branes can
be independently controlled at the level of the worldsheet theory.
This is possible whenever the two-dimensional $\s$ model that
describes the string dynamics is an exact and solvable CFT. The
case that we will be dealing with falls in this class and provides
the first example where the little string holographic duality is
checked with accuracy.

\no Exact conformal $\s$ models serve to generate continuous
families of exact string vacua using e.g. integrable marginal
worldsheet operators. The expected interplay between perturbed
worldsheet $\s$ models and deformed target-space distributions of
source branes has been analyzed in several instances
\cite{Kiritsis:2002xr}-\cite{Petropoulos:2004ir}
%\cite{Kiritsis:2002xr,BFPR,Israel:2003ry,Kiritsis:2003cx,Israel:2004ir,Petropoulos:2004ir}
based on a general criterion established in \cite{CS} for
current-current perturbations. However, the first example of a
clear relationship between a marginal worldsheet operator and a
geometrical deformation of the NS5-brane density distribution has
been worked out in \cite{Marios Petropoulos:2005wu}. There, it was
shown that a continuous deformation of the \emph{circular}
NS5-brane distribution into an \emph{elliptic} one was driven by a
\emph{marginal perturbation} of the ${SU(2)}/{U(1)} \times
{SL(2,\mathbb{R})}/{U(1)}$ worldsheet $\s$ model. The bonus of
this analysis was to demonstrate that the compact parafermions of
the ${SU(2)}/{U(1)}$ theory could be appropriately dressed by
non-compact vertex operators of the ${SL(2,\mathbb{R})}/{U(1)}$ coset and form
a novel kind of operator with anomalous dimension
two which is \emph{non-factorizable} in terms of holomorphic and
anti-holomorphic currents. This marginal operator was responsible
for the circular distribution of NS5-branes being deformed into an
elliptical one.

\no
The deformation of the circle into an ellipsis may be thought of
as one particular mode among an infinitude consisting of battered
circles with $n\in \mathbb{N}$ bumps, distributed with
$\mathbb{Z}_n$ symmetry around the original circle.
Correspondingly, one may expect other combinations of compact
parafermions with non-compact or even compact dressings to provide new
dimension-two operators, each one triggering a mode with a given
number of bumps in the associated NS5-brane distribution. Another
motivation for our article is to demonstrate this
statement for the supersymmetric ${SU(2)}/{U(1)} \times
{SL(2,\mathbb{R})}/{U(1)}$ coset model. Not only this
achievement is interesting per se as an original
conformal-field-theoretical result, but it also establishes one of
the sides of the sought after holographic correspondence that we
advertised previously.

\no
The strategy we will follow is:
\begin{itemize}
  \item First, we analyze in detail the spectrum of conformal
  operators of the unperturbed theory. The latter is the
  near-horizon
  background created by $k$ parallel NS5-branes uniformly distributed on a
  circle which, after T-duality, is the product of two
  supersymmetric Kazama--Suzuki cosets, ${SU(2)}/{U(1)} \times
  {SL(2,\mathbb{R})}/{U(1)}$, as mentioned above. These models have chiral and
  anti-chiral parafermions which are not currents since their
  conformal weights are smaller than one in the case of ${SU(2)}/{U(1)}$ or
  greater than one in the case of $ {SL(2,\mathbb{R})}/{U(1)}$.
  Nonetheless, compact parafermions can be successfully combined
  with non-compact and compact primaries to deliver dimension-two operators.
  For our analysis their  necessary  semiclassical expressions can be worked out
  using group-theory methods.
  \item Next, we move to the LST side and consider a class of
  worldvolume operators whose VEVs describe
  displacements
  of the NS5-branes. These operators must correspond to marginal
  worldsheet operators preserving the original $\mathcal{N}=4$
  superconformal symmetry, like any transverse-space distribution
  of NS5-branes. Using the holographic dictionary, we can indeed associate to
  those LST operators the  marginal operators of the ${SU(2)}/{U(1)} \times
  {SL(2,\mathbb{R})}/{U(1)}$, built previously as compact parafermions appropriately dressed
  with conformal primaries from the non-compact as well as the compact coset in general.
  \item Finally, we can independently check that the worldsheet
  marginal operators constructed by following the holographic
  recipe, do trigger the expected geometric displacements.
  Put differently, we must reinterpret the effect of these
  operators on the $\mathcal{N}=1$ supersymmetric $\s$-model background
  fields and check that these perturbed background fields (metric,
  spin connection, curvature two-form, antisymmetric tensor and
  dilaton) are indeed generated by a distribution of NS5-branes in
  conformity with the distribution predicted by the original LST pattern.
To perform this comparison a delicate interplay between CFT operators and their
semiclassical expressions takes place.
\end{itemize}
These three steps are taken in Secs. \ref{sec:step1},
\ref{sec:step2} and \ref{sec:step3}, respectively. Put together,
they demonstrate the validity of the holographic dictionary and
establish the correspondence among marginal operators of the
supersymmetric ${SU(2)}/{U(1)} \times {SL(2,\mathbb{R})}/{U(1)}$
model and $n$-bump deformations of the circular distribution of
NS5-branes. The main text is followed by several appendices, which
provide the reader with all necessary computational details:
compact and non-compact parafermionic fields and their operator
product expansions (OPEs),
 $\mathcal{N}=4$ and $\cN=2$ extended superconformal algebras,
and finally general properties of the
$n$-bump-deformed geometries (coframes, spin connections and
curvature two-forms).

\section{Neveu--Schwarz five-branes and exact conformal field theories}\label{sec:step1}

In this section we first recall a few facts on the exact CFT
description of the NS5-branes on a point or distributed uniformly
over the circumference of a circle, which concerns, respectively,
the $SU(2)\times \mathbb{R}_\phi$ or $SU(2)/U(1) \times
SL(2,\mathbb{R})/U(1)$ theories and the associated operators. Next
we review material on the classical parafermions relevant for our
paper and develop the semiclassical correspondence of CFT primary
operators to explicit expressions in terms of target-space fields,
which is crucial for the comparisons that we will perform in Sec.
\ref{sec:step3}.

\subsection{Neveu--Schwarz five-branes on a point and on a circle}\label{sec:step11}

A distribution of a large number $k$ of parallel NS5-branes with
density $\rho(\bf x)$ in the transverse $\mathbb{R}^4$ space is
described, to leading order in $\alpha'$, as a supergravity
background specified by a ten-dimensional metric of the form
\begin{equation}
ds^2 =  \eta_{\mu \nu} dx^\mu dx^\nu +  H ({\bf x} ) \delta_{ij}
dx^i dx^j\ ,
\label{met1}
\end{equation}
where $\eta_{\mu\nu}$ is the Minkowski metric on the flat
worldvolume of the NS5-branes parameterized by
$x^\mu,\mu=0,1,\ldots,5$ and ${\bf x}=
\left\{x^i,i=6,7,8,9\right\}$ labels the space $\mathbb{R}^4$
transverse to the NS5-branes. The geometry is accompanied by a
three-form NS--NS field
\begin{equation}
H_{ijk}=\epsilon_{ijk}^{\hphantom{ijk}l} \partial_l H\ ,
\label{hmnr}
\end{equation}
where the indices are lowered and raised with the flat metric of
$\mathbb{R}^4$, and by a dilaton field given by
\begin{equation}
e^{2 (\Phi-\Phi_0) }=H \ ,
\label{dill}
\end{equation}
where $\Phi_0$ is related to the asymptotic string coupling
$g_\mathrm{s}=\exp {2 \Phi_0}$ far from the NS5-branes.

\no
In general, the above background fields provide a solution of
the supergravity equations of motion, which preserves one half of
the maximum supersymmetry, if the function $H({\bf x})$ is
harmonic in $\mathbb{R}^4$. The function $H({\bf x})$ is specified in terms of
the density as
\begin{equation}
\label{haahm}
H({\bf x}) = 1+ \alpha 'k \int_{\mathbb{R}^4} d^4 x'
\frac{\rho({\bf x'})}{|{\bf x}-{\bf x'}|^2}\ .
\end{equation}
Although it is believed that the above background can be promoted
to a string solution valid to all orders in $\alpha'$ for an
arbitrary distribution of NS5-branes, the underlying exact
conformal field theory is not known in general. There are,
however, two special configurations whose near-horizon limit,
corresponding to the harmonic function (\ref{haahm}) with the 1
removed, admit a CFT description. The first is the case of
NS5-branes put at the same point ${\bf x}={\bf 0}$
\cite{Callan:1991dj}. The corresponding harmonic function is
$H({\bf x)}=k/r^2$, with $r$ being the radial distance in
$\mathbb{R}^4$ and after a reparameterization $r = \sqrt{\alpha '
k} \exp \left({\Phi_0} + \nicefrac{\phi}{\sqrt{\alpha'k}}\right)$,
the metric and the three-form become
\begin{equation}
ds^2 =  ds^2 \left(E^{(1,5)}\right) +  d\phi^2 + \alpha ' k\
d\Omega_3^2 \ , \qq H= 2 {\rm Vol}_{S^3}\ ,
\end{equation}
where $d\Omega_3^2$ is the line element of the transverse $S^3$
and ${\rm Vol}_{S^3}$ is its volume form. The dilaton is linear in
$\phi$
\begin{equation}
\Phi=-\frac{q}{2}\phi\ ,\qquad  q=\frac{2}{\sqrt{\alpha' k}} \ .
\end{equation}

\no
The three-sphere of radius $\sqrt{\alpha' k}$  along with the
NS--NS flux can be described by an $SU(2)$  Wess--Zumino--Witten
(WZW) model at level $k$, while the linear dilaton corresponds to
a free boson with background charge $q$. Hence, the near-horizon
region of a system of parallel and coincident NS5-branes admits an
exact conformal field theory description in terms of the
Callan--Harvey--Strominger (CHS) background
\begin{equation}
\mathbb{R}^{5,1} \times \mathbb{R}_{\phi} \times
SU(2)_k\ .\label{pointcft}
\end{equation}
The supersymmetric $SU(2)_k$ WZW model consists of a bosonic
$SU(2)$ WZW model at level $k-2$, whose affine primaries
$\Phi^{su}_{j;m,\bar m}$ have conformal weight
\begin{equation}
h=\frac{j(j+1)}{k}\ ,
\end{equation}
and three free fermions $\psi_a, a=1,2,3$ transforming in the
adjoint representation of $SU(2)$. The conformal primaries of
$\mathbb{R}_\phi$ are $e^{a \phi}$ and their dimension is
\begin{equation}
h=-\frac{1}{2} a (a+q)\ .
\end{equation}
We have also the worldsheet superpartners of $x^\mu$ and $\phi$
given by free fermions $\psi_\mu$ and $\psi_\phi$ . The background
$\mathbb{R}_{\phi} \times SU(2)_k$ supports the small ${\cal N}=4$
superconformal algebra \cite{Kounnas:1990ud} and in App. \ref{CHS}
we present for reference the relevant details. The central charges
of the three CFT factors are
\begin{equation}
%\label{}
\begin{array}{rcl}
c_{5,1} &=& \displaystyle{6 + \frac{6}{2}}\ ,\crbig c_{\phi}
&=&\displaystyle{1+\frac{3 \alpha'}{2} q^2 +\frac{1}{2}}\ ,\crbig
c_k&=&\displaystyle{\frac{3(k-2)}{k}+\frac{3}{2}}\ .
\end{array}
\end{equation}
They add up to $c=15$, as they should in order to have a vanishing
total conformal anomaly. In the rest of this paper $\a'$ is set to
2.

\no The conformal field theory background (\ref{pointcft}) suffers
from a singularity at $\phi=-\infty$ where the string coupling
diverges. Perturbation theory breaks down in this region and hence
(\ref{pointcft}) is a good description of the physics only far
from the NS5-branes. The strong-coupling singularity is due to the
fact that the NS5-branes are coincident since a single NS5-brane
does not develop the ``throat" geometry that results in the linear
dilaton. Hence, separating the NS5-branes should cure the
strong-coupling singularity. At the same time, however, we would
like to keep the benefits of an exact conformal field theory
description. The only known configuration that achieves that, is a
continuous and uniform distribution of NS5-branes on the
circumference of a circle found in \cite{Sfetsos:1998xd}, where it
was shown that in this case the non-trivial part of  the CHS
background (\ref{pointcft})  is replaced, after an appropriate
T-duality,
 by the product of two Kazama--Suzuki coset models
\begin{equation}
\frac{SU(2)_k}{U(1)} \times \frac{SL(2,\mathbb{R})_k}{U(1)}\ ,
\label{cftcircle}
\end{equation}
orbifolded under a $\mathbb{Z}_k$ discrete symmetry.
Actually, as we will see soon, (\ref{cftcircle}) arises as an
exactly marginal deformation of the CHS background.

\subsection{Vertex operators for the coset theories}\label{sec:step12}

The first factor in (\ref{cftcircle}) is the ${\cal N}=2$ minimal
model at level $k$. We will denote its NS--NS sector primaries by
$V^{ su}_{j;m,\bar m}$. Their conformal weight and $R$-charge in
the holomorphic sector are
\begin{equation}
h=\frac{j(j+1)-m^2}{k}\ ,\qq  {\cal Q}_{\mathrm{R}}=-\frac{2m}{k}\
. \label{su2pridata}
\end{equation}
Similar formulas apply for the antiholomorphic sector with $m$
replaced by $\bar m$. A very useful representation of the minimal
model is in terms of a bosonic coset $SU(2)_{k-2}/U(1)$, i.e.~the
compact parafermion theory, and a compact canonically normalized
free boson $P$ \cite{Qiu:1986zf}. The latter bosonizes the two
free fermions that, along with the bosonic $SU(2)_{k-2}/U(1)$
coset, realize the supersymmetric $SU(2)_k/U(1)$ Kazama--Suzuki
model. The operator $V^{ su}_{j;m,\bar m}$ is decomposed as
\begin{equation}
V^{ su}_{j;m,\bar m} = \psi_{j;m,\bar m} \exp\left({i
\frac{2m}{\sqrt{k(k-2)}} P_{\mathrm{L}} + i \frac{2\bar
m}{\sqrt{k(k-2)}} P_{\mathrm{R}}}\right)\ ,
\end{equation}
where $\psi_{j;m,\bar m}$ are primaries of the parafermion theory
\cite{Fateev:1985mm} at level $k-2$ and $P_{\mathrm{L}}(z),
P_{\mathrm{R}}(\bar z)$ are the holomorphic and antiholomorphic
parts of $P$. The conformal dimension of $\psi_{j;m,\bar m}$ is
\begin{equation}
h=\frac{j(j+1)}{k}-\frac{m^2}{k-2}
\label{h16}
\end{equation}
and upon adding to it $\frac{2m^2}{k(k-2)}$, i.e.~the conformal
dimension of the exponential, we obtain the superconformal weight
in (\ref{su2pridata}). An interesting property of the parafermion
theory that we will use is the equivalence of
primaries \cite{Fateev:1985mm}
\begin{equation}
\psi_{j;m,\bar m} \equiv
\psi_{\frac{k-2}{2}-j;-\frac{k-2}{2}+m,-\frac{k-2}{2}+\bar m}\
.\label{cpeq}
\end{equation}
This equivalence relates primaries with $-j\leqslant m \leqslant j
$ originating from $SU(2)_{k-2}$ affine primaries, to primaries
with $j\leqslant m\leqslant k-2-j$. For the latter the conformal
dimension is given by \eqn{h16} with the term $m-j$ added on the
right-hand side.

\no The second factor in (\ref{cftcircle}) is the Kazama--Suzuki
model based on the $SL(2,\mathbb{R})/U(1)$ non-compact coset. Its
NS--NS primaries $V^{ sl}_{j;m,\bar m}$ have conformal weight and
$R$-charge given by
\begin{equation}
h=\frac{-j(j+1)+m^2}{k}\ , \qq {\cal
Q}_{\mathrm{R}}=\frac{2m}{k}\label{sl2pridata}\ .
\end{equation}
As for the ${\cal N}=2$ minimal model, there is a useful
representation of this Kazama--Suzuki model in terms of the
non-compact parafermion theory $SL(2,\mathbb{R})_{k+2}/U(1)$
\cite{Lykken:1988ut} and a free scalar $Q$. The vertex operators
$V^{ sl}_{j;m,\bar m}$ decompose as
\begin{equation}
V^{ sl}_{j;m,\bar m} = \pi_{j;m,\bar m} \exp\left({i
\frac{2m}{\sqrt{k(k+2)}} Q_{\mathrm{L}} + i \frac{2\bar
m}{\sqrt{k(k+2)}} Q_{\mathrm{R}}}\right)\ ,
\end{equation}
where $\pi_{j;m,\bar m}$ are primaries of the non-compact
parafermion theory at level $k+2$, and $Q_{\mathrm{L}}(z)$ and
$Q_{\mathrm{R}}(\bar z)$ are the holomorphic and antiholomorphic
parts of $Q$. The conformal dimensions of $\pi_{j;m,\bar m}$ read:
\begin{equation}
h=-\frac{j(j+1)}{k}+\frac{m^2}{k+2}
\label{h25}
\end{equation}
and, along with the contribution $\frac{2m^2}{k(k+2)}$ of the
exponential, add up to the superconformal weight given in Eq.
(\ref{sl2pridata}). As in the compact case, there is an
equivalence between non-compact parafermion primaries
\begin{equation}
\pi_{j;m,\bar m} =
\pi_{\frac{k-2}{2}-j;\frac{k+2}{2}+m,\frac{k+2}{2}+\bar m}\ .
\label{euivv}
\end{equation}

\subsection{Semiclassical geometry and parafermions}

The coset  $SU(2)_{k-2}/U(1)$ has natural chirally and
anti-chirally conserved objects $\psi, \psi^\dagger$ and $\bar
\psi, \bar \psi^\dagger$ respectively, known as parafermions, with
conformal dimensions \cite{Fateev:1985mm}
\begin{equation}
h=1-\frac{1}{k-2}\ ,\qquad \bar h = 1 - \frac{1}{k-2}\ .
\end{equation}
The parafermion theory, being a coset CFT, it admits a description
as a gauged WZW model \cite{Karabali:1988au}. Semiclassically, the
latter yields a $\sigma$ model with a bell-like target-space
geometry  \cite{Bardacki:1990wj} and a varying dilaton
\cite{Witten:1991yr}
\begin{equation}
ds^2_{SU(2)/U(1)}=k\left(d\theta^2+\tan^2\theta\
d\varphi^2\right)\ , \qquad e^{-2 \Phi} = \cos^2\theta\ ,
\label{compaco}
\end{equation}
with $\theta \in [0,\pi], \varphi \in [0,2\pi)$ and $\varphi
\equiv \varphi + \frac{2 \pi}{k}$. In the standard parafermion
theory the compact scalar $\varphi$ has period $2 \pi$ but here we
have changed its period to $2\pi/k$ so that it corresponds to the
$\mathbb{Z}_k$ orbifold of the original parafermion theory. In
terms of the $\sigma$-model variables, the classical parafermion
fields read \cite{Bardacki:1990wj}:
\begin{equation}
\label{classcp} \psi =\big(\partial \theta - i \tan\theta\
\partial \varphi\big) e^{-i(\varphi+\phi_1)}\ ,\qquad  \psi^\dagger
=\big(\partial \theta + i \tan\theta\ \partial \varphi\big)
e^{i(\varphi+\phi_1)} \
\end{equation}
and
\begin{equation}
\label{classcp1} \bar \psi = \big(\bar \partial \theta - i
\tan\theta\  \bar \partial \varphi\big) e^{-i(\varphi-\phi_1)}\ ,
\qquad  \bar \psi^\dagger = \big(\bar
\partial \theta + i \tan\theta\ \bar\partial \varphi\big)
e^{i(\varphi-\phi_1)}\ .
\end{equation}
The overall normalization is chosen such that the OPE of the
corresponding quantum parafermions in App. \ref{pOPE} agree with
the Poisson brackets of their above classical counterparts. The
parafermions have their origin in the currents $J^\pm$ and $\bar
J^\pm $ of the $SU(2)_k$ theory. In the gauged theory they are
dressed with gauge fields that render them gauge-invariant. That
explains also the presence of the phase $\phi_1$ which is a
non-local function of the variables $\th$ and $\varphi$. Its
explicit expression is not needed here (see, for instance,
\cite{Marios Petropoulos:2005wu}), but it is necessary for
ensuring on-shell conservation of the parafermions
\begin{equation}
\label{dik1}
\bar \partial \psi = \bar \partial \psi^\dagger =0\ , \qq
\partial  \bar \psi = \partial  \bar\psi^\dagger =0\ .
\end{equation}
The non-local phase $\phi_1$ should drop out in expressions having a clear local field theory
interpetation, for instance those appearing in the two-dimensional $\s$ model actions as we shall see.

\no The non-compact coset  $SL(2,\mathbb{R})_{k+2}/U(1)$ has also
natural chirally and anti-chirally conserved objects $\pi,
\pi^\dagger$ and $\bar \pi, \bar \pi^\dagger$ respectively, known
as non-compact parafermions, with conformal dimensions
\cite{Lykken:1988ut}
\begin{equation}
h=1+\frac{1}{k+2}\ ,\qquad \bar h = 1 + \frac{1}{k+2}\ .
\end{equation}
The non-compact-parafermion theory admits a semiclassical
description in terms of a $\sigma$ model with either a
cigar-shaped or a trumpet-shaped geometry along with a non-trivial
dilaton \cite{Witten:1991yr}. We will consider the trumpet
picture, specified by
\begin{equation}
ds^2_{SL(2,\mathbb{R})/U(1)} = k\left(d\rho^2 + \coth^2 \rho\
d\omega^2\right)\ , \qq e^{-2 \Phi} = \sinh^2 \rho\ ,
\label{companonco}
\end{equation}
with coordinates $\rho \in [0,\infty), \omega \in [0,2\pi)$ and
$\omega \equiv\omega+ \frac{2 \pi}{k}$. This metric is singular near
$\rho=0$ but its T-dual, namely the cigar, is well-defined and
provides an equivalent (up to T-duality) semiclassical description
of the non-compact parafermion theory. The classical non-compact
parafermion fields are
\begin{eqnarray}\
\pi = \big(\partial \rho +i \coth\rho\ \partial \omega\big)
e^{i(\omega+\phi_2)}\ ,\qq \pi^\dagger = \big(\partial \rho-i
\coth\rho\ \partial \omega\big) e^{-i(\omega+\phi_2)}
\label{nonc1} \ea and \be \bar \pi = \big(\bar \partial \rho +i
\coth\rho\ \bar
\partial \omega\big) e^{i(\omega-\phi_2)}\ ,\qq \bar \pi^\dagger =
\big(\bar
\partial \rho -i \coth\rho\ \bar
\partial \omega\big) e^{-i(\omega-\phi_2)}\ .
\label{nonc2} \ee The phase $\phi_2$ is non-local and ensures
on-shell conservation laws similar to those of the compact case,
Eq. \eqn{dik1}.

\no The full conformal field theory (\ref{cftcircle}) corresponds
semiclassically to a $\sigma$ model with metric and dilaton given
by
\begin{equation}
ds^2=k\left(d\theta^2+\tan^2\theta\ d\varphi^2+d\rho^2 + \coth^2
\rho\ d\omega^2\right)\ , \qq e^{-2 \Phi} = \cos^2\theta \sinh^2
\rho\ . \label{sigmacirclecft}
\end{equation}
The relation of this background to that of NS5-branes distributed uniformly over
a circle, follows explicitly by first changing coordinates as $\varphi=\tau$ and
$\om=\tau + \psi$ and then performing a T-duality transformation with respect to $\tau$ \cite{Sfetsos:1998xd}.
The focal point of this paper will be
the interplay between the description of deformations of the
circular distribution of NS5-branes in the $\sigma$-model language
and the corresponding operators in the exact conformal field
theory description. For that, an important ingredient will be the semiclassical expressions of the
conformal primary fields of the theory (\ref{cftcircle}), to the description of which we now turn.

\subsection{Semiclassical description of conformal primaries}

Conformal primaries in (\ref{cftcircle})  are products of
primaries of each factor. First we consider the semiclassical
description of the primaries of the WZW model for the non-compact
group $SL(2,\mathbb{R})$ from which the semiclassical primaries
for the non-compact coset $SL(2,\mathbb{R})/U(1)$ follow by
rendering them gauge-invariant. These primaries are built up using
the group element\footnote{In our presentation we follow and
extend the brief discussion in \cite{Marios Petropoulos:2005wu}.
For an extensive overview and general expressions see
\cite{vilenkinbook} and also \cite{Chaudhuri:1992yca}, where some
of the primaries have been used in relation to the physics of the
two-dimensional black hole.}
\begin{equation}
  \begin{pmatrix}
 g_{++} & g_{+-}\\
 g_{-+} & g_{--}
  \end{pmatrix}
=
  \begin{pmatrix}
\cosh\r\ e^{i(\th_{\mathrm{L}}+\th_{\mathrm{R}})/2} & \sinh\r\
e^{-i(\th_{\mathrm{L}}-\th_{\mathrm{R}})/2} \cr \sinh\r\
e^{i(\th_{\mathrm{L}}-\th_{\mathrm{R}})/2} & \cosh\r \
e^{-i(\th_{\mathrm{L}}+\th_{\mathrm{R}})/2}
    \end{pmatrix}
\end{equation}
and they transform in the $\left(\ha,\ha\right)$ representation of
$SL(2,\mathbb{R})_{\mathrm{L}}\times
SL(2,\mathbb{R})_{\mathrm{R}}$ with $U(1)$ charges $\left(\pm \ha
,\pm\ha\right)$, in all four combinations, in accordance with
their index. The explicit transformation rules referring to
$SL(2,\mathbb{R})_{\mathrm{L}}$ are
\begin{equation}
\label{trdf}
\begin{array}{rclrcl}
\d_0 g_{\pm\pm}
 &=&\displaystyle{
\mp {i\ov 2} g_{\pm\pm}} \ ,& \qquad \d_0g_{\pm\mp}
&=&\displaystyle{\mp {i\ov 2} g_{\pm\mp} }\ , \crbig \d_- g_{++}
 &=&
i g_{-+} \ ,& \qquad \d_+ g_{++} &=&0\ ,  \crbig
 \d_- g_{+-}
 &=& ig_{--}
\ ,& \qquad \d_+ g_{+-} &=&0\ ,  \crbig
 \d_- g_{-+}
 &=&0
\ ,& \qquad \d_+ g_{-+} &=&-i g_{++}\ ,   \crbig
 \d_- g_{--}
 &=&0
\ ,& \qquad
 \d_+ g_{--}
&=&-i g_{+-}\ ,
\end{array}
\end{equation}
and act only on the first index of the group elements. The similar
transformations with respect to $SL(2,\mathbb{R})_{\mathrm{R}}$
acting on the second index of the group elements, may have the
same or the opposite signs as compared to those in (\ref{trdf}) since the two
transformations are unrelated. We choose the opposite sign since
in the non-compact coset model we will gauge the vectorial $U(1)$
subgroup instead of the axial, as we will shortly discuss.

\no
Being finite-dimensional,
the above is not a unitary representation of
$SL(2,\mathbb{R})_{\mathrm{L}}\times
SL(2,\mathbb{R})_{\mathrm{R}}$, but we may construct other
irreducible representations  that are unitary by appropriate
multiplications and inversions of the above group elements. In
particular, for the positive and negative discrete series, for
given spin $j$, $m$ takes the values $\pm(j+1,j+2,\dots )$. In
order for the semiclassical description to remain valid, we will
assume $j\ll k$. Then it is obvious that the following expressions
for the semiclassical primaries are unique
\begin{equation}
\label{prim1}
\begin{array}{rclrcl}
\pi_{j;j+1,j+1}&=&\displaystyle{{1\ov g_{--}^{2(j+1)}}}\   ,
&\quad \pi_{j;-j-1,-j-1}&=&\displaystyle{{1\ov g_{++}^{2(j+1)}}}\
, \crbig \pi_{j;j+1,-j-1}
 &=&\displaystyle{{1\ov g_{-+}^{2(j+1)}}}\ , &\quad
 \pi_{j;-j-1,j+1}&=&\displaystyle{{1\ov g_{+-}^{2(j+1)}}}\ ,
\end{array}
\end{equation}
since they represent highest- or lowest-weight states. The other
members of the representation are obtained by transforming
appropriately the states in \eqn{prim1} with $\d_\pm$ using
\eqn{trdf}. It is crucial for the precise comparison that we
perform in the next section to have an agreement between
normalization factors of various operators in the semiclassical
and the exact CFT approaches. Hence we will normalize them
according to (\ref{ncparapriOPE}) as
 \be \pi_{j;m\pm 1,\bar m} =
 {1\ov m\pm (j+1)}\ \d_\pm \p_{j;m,\bar m}\ ,
 \ee
for the left as well as for the right $SU(2)$ transformations.
Here we present only the expressions one needs in this paper \be
\pi_{j;j+2,j+1}=i {g_{+-}\ov g_{--}^{2 j+3}}\ ,\ \
\pi_{j;j+1,j+2}=-i{g_{-+}\ov g_{--}^{2 j+3}}\ , \ \
\pi_{j;j+2,j+2}={2 (j+1) g_{+-}g_{-+}-1\ov 2 (j+1)g_{--}^{2 j+4}}\
. \label{prim2} \ee

\no For the parafermionic coset theory, corresponding to the
gauging of $SL(2,\mathbb{R})$ with respect to the vector $U(1)$
subgroup described semiclassicaly in terms of \eqn{companonco}, we
have the transformation $\d \th_{\mathrm{L}} = -\e $ and $\d
\th_{\mathrm{R}} = \e$, so that the appropriate unitary gauge
fixing is $\th_{\mathrm{L}}=\th_{\mathrm{R}} =\om$.\footnote{ The
vector transformation is consistent with the sign difference
between the left and right $SU(2)$ transformations, mentioned
below \eqn{trdf}. Indeed, since
$i(\th_{\mathrm{L}}+\th_{\mathrm{R}})=\ln(g_{++}/g_{--})$ and
$i(\th_{\mathrm{L}}-\th_{\mathrm{R}})=\ln(g_{-+}/g_{+-})$, we can
easily see that for the left (right) transformations with the
$U(1)$ subgroup one obtains $\d\th_{\mathrm{R}}=\e_{\mathrm{L}}$
and $\d\th_{\mathrm{L}}=0$ ($\d\th_{\mathrm{R}}=0$ and
$\d\th_{\mathrm{L}}=\e_{\mathrm{R}}$). Since by definition a
vector transformation has $\e_{\mathrm{L}}=-\e_{\mathrm{R}}$, we
see that this is indeed consistent with the transformations
$\d\th_{\mathrm{L}}=-\e=-\d\th_{\mathrm{R}}$.} The elements
$g_{\pm\pm}$ are gauge-invariant, whereas $g_{\pm\mp}$ are not.
Those become gauge-invariant provided they are multiplied by the
non-local phase factor $\phi_2$ that appears in the non-compact
parafermions. The gauge-invariant group elements that should be
used in \eqn{prim1} and \eqn{prim2} to construct the expressions
for the semiclassical primaries of the parafermionic theory are
thus
\begin{equation}
g_{\pm\pm}|_{\rm g.-inv.}=\cosh\rho\ e^{\pm i \omega}\ ,\qq
g_{\pm\mp}|_{\rm g.-inv.}=\sinh\rho\ e^{\pm i \phi_2}\ .
\label{ginva}
\end{equation}
Form now on we drop the indicated index, keeping in mind that the associated semiclassical primaries
correspond to the parafermionic coset theory.
In the semiclassical correspondence that we will establish we will take into account the leading
$1/k$-correction to their classical dimension which is zero.
Therefore the dimension of the above semiclassical primaries of the parafermionic theory \eqn{h25}, becomes
\be
h_{j; j + \ell} = h_{j; -j - \ell}= {(2\ell-1)j+\ell^2\ov k} + {\cal O}(1/k^2)\ ,
\ee
accordingly for the left or the right factor and where $\ell=1,2$ in our case.

\no
For the case of the semiclassical primaries of the compact coset
$SU(2)/U(1)$ the procedure is quite similar. We will be brief since the expressions we
need in this paper are fewer than those needed from the non-compact coset.
The semiclassical $SU(2)$ compact primaries are built up using the group element (we will use tildes so that
there is no confusion with the $SL(2,\mathbb{R})$ group element we used above)
\begin{equation}
  \begin{pmatrix}
    \tilde g_{++}  & \tilde g_{+-} \\
    \tilde g_{-+} & \tilde g_{--}
  \end{pmatrix}
=
  \begin{pmatrix}
    \cos\th\ e^{i(\th_{\mathrm{L}}+\th_{\mathrm{R}})/2} & \sin\th\ e^{-i(\th_{\mathrm{L}}-\th_{\mathrm{R}})/2} \\
    -\sin\th\ e^{i(\th_{\mathrm{L}}-\th_{\mathrm{R}})/2} & \cos\th\
e^{-i(\th_{\mathrm{L}}+\th_{\mathrm{R}})/2}
  \end{pmatrix} \ .
\end{equation}
These transform similarly to \eqn{trdf} and form the
$\left(\ha,\ha\right)$ unitary representation of
$SU(2)_{\mathrm{L}}\times SU(2)_{\mathrm{R}}$ with $U(1)$ charges
$\left(\pm \ha ,\pm\ha\right)$, in all four combinations and in
accordance with their index. In the present paper we will need
only the expressions for
 \be
 \psi_{j;j,j}=\tilde g_{++}^{2j}\ ,\qq \psi_{j;-j,-j}=\tilde
 g_{--}^{2j}\ , \qq \psi_{j;j,-j}=\tilde g_{+-}^{2j}\ ,\qq
 \psi_{j;-j,j}=\tilde g_{-+}^{2j}\ ,
  \ee
which, for fixed $j$, are highest- or lowest-weight
representations for the left and right $SU(2)$ factors. For the
parafermionic $SU(2)/U(1)$ coset theory with axial gauging,
corresponding to the background \eqn{compaco}, we have the
transformation $\d \th_{\mathrm{L}} = \e $ and $\d
\th_{\mathrm{R}} = \e$, so that the appropriate unitary gauge
fixing is $\th_{\mathrm{L}}=-\th_{\mathrm{R}} =\varphi$. The
elements $g_{\pm\mp}$ are gauge-invariant, whereas $g_{\pm\pm}$
are not. As before, it turns out that they become gauge-invariant
when they are multiplied by the non-local phase factor that
appears in the compact parafermions. Hence
\begin{equation}
\tilde g_{\pm\mp}|_{\rm g.-inv.}=\pm \sin\th\ e^{\mp i \varphi}\ ,
\qq \tilde g_{\pm\pm}|_{\rm g.-inv.}=\cos\th\ e^{\mp i \phi_1}\ .
\end{equation}
The dimensions of the above semiclassical primaries \eqn{h16} in
the parafermionic theory, up to the order we are interested in,
are
 \be h_{j; j}=h_{j;- j}= {j\ov k} + {\cal O}\left(1/k^2\right)\
 , \ee accordingly for the left or the right factor.

\section{Holographic approach to NS5-brane deformations}\label{sec:step2}

In this section, after reviewing the procedure by which the theory
is deformed based on the holographic conjecture, we compute
explicitly the operators corresponding to our cases. Subsequently
we find their semiclassical expressions in terms of target-space
fields. Then we specialize our findings to some simple cases like
the deformation of a point distribution into a circular one and
that of a circular into one of a different radius. We then pay
particular attention to the deformation of a circular into a an
elliptical distribution which captures most of the essential points
of our construction. Finally, we apply our results to the case of
a general distribution which,  compared to the elliptical one,
presents some new features such as the appearance of composite
operators.

\subsection{Holographic dictionary: generalities}

 An interesting feature of both (\ref{pointcft}) and (\ref{cftcircle})
is that asymptotically (i.e.~large $\rho$ for  (\ref{cftcircle})
in terms of its $\sigma$-model description
(\ref{sigmacirclecft})), they are linear dilaton space--times. It
has been proposed that string theory on such space--times provides
a holographic description of the mysterious non-gravitational
string theory, known as little string theory, that lives on the
worldvolume of NS5-branes in the decoupling limit where the
asymptotic string coupling is taken to zero \cite{Aharony:1998ub}.
This correspondence is very similar to the usual AdS/CFT duality
since it relates the decoupled theory on a stack of branes with
the supergravity or string theory on the near-horizon geometry
induced by the branes \cite{Maldacena:1997re, gubser, witten}.

\no The spectrum of states in backgrounds that asymptote to a
linear dilaton falls into three classes with distinct physical
significance. For instance in the CHS background, there are
delta-function normalizable states whose vertex operators behave
for large $\phi$ as $e^{\left(-\frac{q}{2}+i \lambda\right)\phi}$
with real $\lambda$. They describe  incoming and outcoming waves
carrying momentum $\lambda$ along the holographic direction
$\phi$. Besides these states, the theory contains also
normalizable states that decay rapidly as $\phi \rightarrow
\infty$. Hence, they are supported in the strong-coupling region
of large negative $\phi$ and they can be thought of as bound
states associated
 with the NS5-branes. Finally, there exist non-normalizable states
whose wavefunctions diverge at the weakly-coupled boundary $\phi \rightarrow \infty$.

\no The holographic duality conjectures a correspondence between
vertex operators of the string theory on the asymptotic linear
dilaton background and deformations of the dual little string
theory \cite{Aharony:2003vk,Aharony:2004xn}. More precisely,
adding to the worldsheet Lagrangian a non-normalizable operator
$V_{\rm non-nor.}$, corresponds to perturbing the Lagrangian of
the dual theory with an appropriate dual gauge-invariant operator
$W_V$. If, instead, we add to the worldsheet theory the
normalizable version $V_{\rm nor.}$  of the same vertex operator,
the dual theory does not change
 but the dual operator $W_V$ acquires a VEV $\langle W_V\rangle$  \cite{klewit}.
Since the geometry of the NS5-branes and their deformations are
encoded in the VEVs of the adjoint scalar fields living on their
worldvolume, we see that by employing the holographic
correspondence  we can uncover the associated deformations of the
underlying (dual) CFT.

\no In the little string theory side, a basic class of operators
we would like to consider and which encode all the information on
the arrangement of the NS5-branes in their transverse
$\mathbb{R}^4$ is given by chiral and gauge-invariant combinations
of the adjoint scalar fields $\Phi^i, \; i=6,7,8,9$. The
eigenvalues of these fields parameterize the positions of the
NS5-branes in the four transverse directions, i.e.~the moduli
space of vacua of the little string theory. The operators of
interest are $ {\rm tr} \left(\Phi^{i_1}\Phi^{i_2}\cdots
\Phi^{i_{2j+2}}\right)$ with $2j=0,1,\ldots,(k-2)$ and where we
keep only the symmetric and traceless components in the indices
$(i_1,i_2,\ldots,i_{2j+2})$ so that the operator is in a short
representation of the supersymmetry algebra.\footnote{There is
also a subtlety pertaining to the precise definition of the trace.
In principle, one should consider the usual single-trace along
with multi-trace operators. However, for $j\ll k$, which will be the
regime of our interest, the  multi-trace contributions will be
negligible and it will suffice to consider the single-trace ones
\cite{Aharony:2003vk,Aharony:2004xn}.}

\no
 The dictionary established in \cite{Aharony:2003vk,Aharony:2004xn} is
 \begin{equation}\label{holodic}
{\rm tr} \left(\Phi^{i_1}\Phi^{i_2}\cdots\Phi^{i_{2j+2}}\right)
\leftrightarrow e^{-\varphi-\bar\varphi} \left(\psi\bar \psi
\Phi_j^{{ su}}\right)_{j+1;m,\bar m} e^{-q(j+1)\phi}\ ,
\end{equation}
where the right-hand side refers to operators in the CHS background. We use
the normalizable version of the CFT operators since we are interested in
describing VEVs in the little string theory. We denoted by $\varphi,\bar\varphi$ the bosonized
 superconformal ghosts (which should not be confused with the compact coordinate
 of the  bell geometry (\ref{compaco})),
 $\Phi^{{ su}}_j$ is an affine primary of the
 bosonic $SU(2)_{k-2}$ WZW model and the notation
 $\left(\psi\bar \psi \Phi_j^{{ su}}\right)_{j+1;m,\bar m}$ means that we should couple
 the fermions $\psi^a, a =3,\pm$ in the adjoint of $SU(2)$
with the bosonic primary in a primary of total spin $j+1$ and
$\left(J_3^{\rm tot},\bar J_3^{\rm tot}\right)=(m,\bar m)$. We
refer the reader to the App. \ref{CHS} for further details on the
notation. The values of $m$ and $\bar m$ are determined by the
indices appearing at the left.

\no When the operators  $ {\rm tr}
\left(\Phi^{i_1}\Phi^{i_2}\cdots \Phi^{i_{2j+2}}\right)$ acquire
non-zero VEVs, the Lagrangian of the dual worldsheet conformal
field theory is perturbed as
\begin{equation}
{\cal L}={\cal L}_0+\Big(\lambda_{j;m,\bar m} G_{-\frac{1}{2}}\bar G_{-\frac{1}{2}}
 \left(\psi\bar \psi \Phi_j^{{ su}}\right)_{j+1;m,\bar m}
e^{-q(j+1)\phi} + {\rm c.c.}\Big)\ ,
\label{N=4pert}
\end{equation}
where we have omitted the bosonized ghosts since  we will be
working in the $0$-picture. These worldsheet deformations are
marginal since $ \left(\psi\bar \psi \Phi_j^{{
su}}\right)_{j+1;m,\bar m} e^{-q(j+1)\phi}$ has conformal weights
$(h,\bar h)=(1/2,1/2)$. Furthermore, they should leave unbroken
the ${\cal N}=4$ superconformal symmetry of the original
worldsheet theory since any configuration of parallel NS5-branes
with arbitrary transverse positions preserves one-half of the
maximum space--time supersymmetry. We check that ${\cal N}=4$ is
indeed preserved in App. \ref{CHS}. The couplings
$\lambda_{j;m,\bar m}$ are specified in terms of the VEVs of the
LST operators while the supersymmetry generators
$G_{-\frac{1}{2}}$ and $\bar G_{-\frac{1}{2}}$ correspond to the
supercurrent $G$  defined in App. \ref{CHS}. Notice that the
perturbations we add in the worldsheet theory dominate at the
region of strong coupling $\phi \rightarrow -\infty$ and provide a
worldsheet potential that regularizes the strong-coupling
singularity.

\no We will be interested in planar configurations of NS5-branes,
i.e.~distributions on the transverse plane $x^8-x^9$. It is very
convenient then, following \cite{Aharony:2003vk,Aharony:2004xn},
to use a parameterization of the moduli space in terms of two
complex variables that span the two orthogonal hyperplanes
transverse to the NS5-branes:
\begin{equation}\label{defAB}
A\equiv \Phi^6+i \Phi^7\ , \qq
B\equiv \Phi^8+i \Phi^9\ .
\end{equation}
Embedding the rotational $SO(2)_A \times SO(2)_B$ of the $A$ and
$B$ planes in the $SU(2)_{\mathrm{L}} \times SU(2)_{\mathrm{R}}$
symmetry of the CHS background  so that $SO(2)_A$ is generated by
$J_3^{\rm tot}-\bar J_3^{\rm tot}$ and $SO(2)_B$ is generated by
$J_3^{\rm tot}+\bar J_3^{\rm tot}$, yields the following charge
assignments
\begin{equation}
m_A=\frac{1}{2}\ ,\quad \bar m_A=-\frac{1}{2} \ ,\quad m_B=\frac{1}{2}\ ,\quad \bar m_B=\frac{1}{2}
\ .
\end{equation}
Combining those with the general relation (\ref{holodic}) leads to
the  following correspondences
 \be {\rm  tr} \left(A^l
B^{2j+2-l}\right)\leftrightarrow e^{-\varphi-\bar\varphi}
\left(\psi\bar \psi\Phi_j^{{ su}}\right)_{j+1;j+1,j+1-l}
e^{-q(j+1)\phi}\ \label{cor1point} \ee and \be {\rm tr} \left(A^l
(B^*)^{2j+2-l}\right) \leftrightarrow e^{-\varphi-\bar\varphi}
\left(\psi\bar \psi\Phi_j^{{ su}}\right)_{j+1;-j-1+l,-j-1}
e^{-q(j+1)\phi}\label{cor2point}\ . \ee

\no We set $\langle A \rangle=0$ since we will study  NS5-branes
distributed on the  $B$ plane and fixed at $x^6=x^7=0$. Their
positions are parameterized by $k$ complex numbers $b_n,\;
n=1,2,\ldots,k$
\begin{eqnarray}\label{vevAB}
\langle B\rangle&=&{\rm diag}\left(b_1,b_2,\ldots,b_k\right)\ ,
\qq \sum_{n=1}^k b_n=0\ ,
\end{eqnarray}
where the condition on their sum
ensures that the center of mass of the NS5-brane system does not change
in accordance with the fact that the corresponding $U(1)$ degree of freedom
is not part of the interacting LST. Since $\langle A \rangle=0$ only the operators with $l=0$
 from (\ref{cor1point}) and (\ref{cor2point}) have non-vanishing VEVs. Hence, the
 holographic dictionary becomes
 \begin{equation}
{\rm  tr} (B^{2j+2}) \leftrightarrow e^{-\varphi-\bar\varphi}
\psi^+ \bar \psi^+ \Phi^{{ su}}_{j;j,j} e^{-q(j+1)\phi}\
,\label{pointcftchiral}
\end{equation}
since $\psi^+, \bar \psi^+$ have $\left(J_3^{\rm tot},\bar
J_3^{\rm tot}\right)=(1,1)$.

\no
A different representation of these operators comes from
the decomposition
\begin{equation}\label{decomp}
\mathbb{R}_{\phi}\times SU(2)_k \equiv \mathbb{R}_\phi \times
\left(U(1)_k \times \frac{SU(2)_k}{U(1)}\right)\bigg/\mathbb{Z}_k\
.
\end{equation}
The infinite cylinder $\mathbb{R}_\phi \times U(1)_k$ is
parameterized by $\phi$ and $Y$ with the latter defined as
\begin{equation}\label{J3}
J_3^{\rm tot}=\frac{i}{q} \partial Y\ .
\end{equation}
The ${\cal N}=2$ minimal model $SU(2)_k/U(1)$ can be described
in terms of a Landau--Ginzburg superfield $\chi$ with superpotential
\begin{equation}\label{LGW1}
W=\chi^k\ .
\end{equation}
Then we can write
\begin{equation}\label{LGW2}
\psi^+\bar \psi^+ \Phi^{{ su}}_{j;j,j} e^{-q (j+1) \phi} =
\chi^{k-2(j+1)}e^{-q(j+1)\Phi}\ ,
\end{equation}
where $\Phi$ is a chiral superfield whose bottom component is
$\phi-iY$ (and, as usual, we will denote both of them by the same
symbol from now on). The perturbed Lagrangian can be written as
\begin{equation}\label{Lpert}
{\cal L}={\cal L}_0+\left(\sum_j \lambda_j \int d^2\theta
\chi^{k-2(j+1)} e^{-q(j+1)\Phi} +{\rm c.c.}\right)\ ,
\end{equation}
where the couplings $\lambda_j$ are specified in terms of the
locations of the NS5-branes on the $B$-plane: $\lambda_j ={1\ov k}
\left\langle {\rm tr} \left(B^{2j+2}\right) \right\rangle$, with
the proportionality factor $1/k$ being included so that the
coupling, for generic NS5-distributions, is appropriately
normalized.

\no
So far the discussion applies to a configuration of NS5-branes located at a point
in $\mathbb{R}^4$, where the dual conformal field theory is (\ref{pointcft}),
and (\ref{holodic}) associates deformations of the NS5-branes around the point
with marginal operators in the CHS background. If, instead, we are
interested in deformations of the circular configuration of NS5-branes,
which admits an exact CFT description in terms of (\ref{cftcircle}), we would like
to associate VEVs of the chiral operators $ {\rm tr} (\Phi^{i_1}\Phi^{i_2}\cdots \Phi^{i_{2j+2}})$
with marginal operators in  (\ref{cftcircle}). A way to find the latter is to consider
 (\ref{cftcircle}) as a deformation of $\mathbb{R}_\phi \times SU(2)_k  \simeq
 \mathbb{R}_\phi \times \Big(U(1)_k \times \frac{SU(2)_k}{U(1)}\Big)\Big/\mathbb{Z}_k$
 with the cylinder $\mathbb{R}_\phi \times U(1)_k$ being deformed to the
 $SL(2,\mathbb{R})_k/U(1)$ coset theory. For the operators of interest the dictionary becomes
 \begin{equation}
 {\rm  tr} \left(B^{2j+2}\right)  \leftrightarrow
 e^{-\varphi-\bar\varphi}
 V^{ su}_{\frac{k}{2}-j-1;-\frac{k}{2}+j+1,-\frac{k}{2}+j+1} V^{ sl}_{j;j+1,j+1}\ .
\label{circlecftchiral}
\end{equation}
The corresponding worldsheet deformations are
\begin{equation}
{\cal L}={\cal L}_0+\Big(\lambda_{j} G_{-\frac{1}{2}}\bar
G_{-\frac{1}{2}}
V^{su}_{\frac{k}{2}-j-1;-\frac{k}{2}+j+1,-\frac{k}{2}+j+1} V^{
sl}_{j;j+1,j+1} + {\rm c.c.}\Big)\label{wdcircle}
\end{equation}
and we can use (\ref{su2pridata}) and (\ref{sl2pridata}) to check that they
are indeed marginal.

\no A very interesting feature of the CFT operators in
(\ref{pointcftchiral}) and (\ref{circlecftchiral}) is that they
are (chiral, chiral) \cite{Lerche:1989uy}. One can verify, using
the formulas in App. \ref{CHS} for (\ref{pointcftchiral}) and in
App. \ref{pOPE} and App. \ref{sucal} for (\ref{circlecftchiral}), that
\begin{equation}
\begin{array}{rcl}
&&\displaystyle{G^+(z) \psi^+ \bar \psi^+ \Phi^{{ su}}_{j;j,j}
e^{-q(j+1)\phi} (w,\bar w) } \sim 0 \ ,  \crbig &&\displaystyle{
G^+(z) V^{ su}_{\frac{k}{2}-j-1;-\frac{k}{2}+j+1,-\frac{k}{2}+j+1}
V^{ sl}_{j;j+1,j+1}(w,\bar w)} \sim 0\ ,
\end{array}
\end{equation}
and similarly for the antiholomorphic sector. Since the conformal
weights of chiral operators are fixed in terms of their
$R$-charges, the corresponding worldsheet deformations are exactly
marginal. It is interesting to notice that spreading the
NS5-branes on the $A$-plane while keeping them at a fixed point in
the $B$-plane, $\langle B \rangle =0$,
 corresponds holographically to CFT operators that are (chiral, antichiral).
Performing, instead, general (non-planar) deformations of the NS5-branes
where both $\langle A \rangle$ and $\langle B \rangle$ are non-zero triggers
non-chiral operators and therefore we lose exact marginality.
Finally, we mention that interesting properties of the chiral ring comprised by
the operators \eqn{circlecftchiral} has been studied in \cite{topol1,topol2}.

\no We proceed now with the computation of the worldsheet
deformation in (\ref{wdcircle}). The supercurrent is
$G=\frac{1}{\sqrt{2}}(G^++G^-)$ and the action of the supercharge
$G_{-\frac{1}{2}}$ on the operator is captured by the simple pole
of its OPE with $G$. As we mentioned above, the operators under
consideration are chiral and hence the singularities in their OPE
with $G$ will come only from the action of $G^-$. Using the
decomposition of the supersymmetric coset primaries presented in
the previous subsection, the decomposition of the supercurrents
from App. \ref{pOPE} and  the parafermionic OPEs from App.
\ref{sucal}, we can derive the following expressions \ba
\label{G-VSU}
 G^{ -su}(z) V^{ su}_{\frac{k}{2}-j-1;-\frac{k}{2}+j+1,-\frac{k}{2}+j+1}(w,\bar w)
&\sim & {\sqrt{2}\ \a_1 \ov z-w}\
 \psi_{\frac{k}{2}-j-1;-\frac{k}{2}+j+2,-\frac{k}{2}+j+1} \nonumber\\
 &\times & \exp\left({i \frac{2(j+1)}{\sqrt{k(k-2)}}P_{\mathrm{L}}+i \frac{2(j+1)-k}{\sqrt{k(k-2)}}P_{\mathrm{R}} }\right)\
 \label{G-VSL}
\ea
and
\ba
 G^{ -sl}(z) V^{ sl}_{j;j+1,j+1}(w,\bar w) &\sim& {\sqrt{2}\ \a_2\ov z-w}
 \ \pi_{j;j+2,j+1} \nonumber\\
 &\times&
  \exp\left({i \frac{2(j+1)-k}{\sqrt{k(k+2)}}Q_{\mathrm{L}}+i \frac{2(j+1)}{\sqrt{k(k+2)}}Q_{\mathrm{R}}}\right)\ ,
 \end{eqnarray}
with coefficients
 \be \a_1 ={1 \ov \sqrt{ k}}
 \left(k-2(j+1)\right)\ ,\qq \a_2 ={2 (j+1)\ov \sqrt{k}}\ . \ee
Similar expressions hold for the actions of the antiholomorphic
supercurrents.

 \no
The operator we add to the worldsheet Lagrangian comes
 from the action of the total supercharges $G^-=G^{ -su}+G^{ -sl}$ and $\bar G^-=
 \bar G^{ -su}+\bar G^{ -sl}$ on
 \begin{equation}\label{ellipsisdef2}
V^{ su}_{\frac{k}{2}-j-1;-\frac{k}{2}+j+1,-\frac{k}{2}+j+1} V^{
sl}_{j;j+1,j+1}: = {\cal V}^{ su} {\cal V}^{ sl}\ ,
\end{equation}
 so that the result, after taking into
account the  factor $1/2$ coming from the definitions of $G$ and
$\bar G$, schematically reads:
 \begin{equation}\label{gd}
 \frac{1}{2}\Big({\cal V}^{ sl} (G^{ -su} \bar G^{ -su} {\cal V}^{ su}) + (G^{ -su} {\cal V}^{ su})
 (\bar G^{ -sl} {\cal V}^{ sl}) + (G^{ -sl} {\cal V}^{ sl})(\bar G^{ -su} {\cal V}^{ su})
 +{\cal V}^{ su} (G^{ -sl} \bar G^{ -sl} {\cal V}^{ sl})\Big).
 \end{equation}
The total deformation of the Lagrangian  is given by (\ref{gd})
multiplied by the couplings $\lambda_j$ and upon adding to it its
complex conjugate so that the total expression is real.
Explicitly, the four operators in (\ref{gd}) read:\begin{eqnarray}
&&C^{(1)}_{j,k}\ \alpha_1^2\
\psi_{\frac{k}{2}-j-1;-\frac{k}{2}+j+2,-\frac{k}{2}+j+2}\
 \pi_{j;j+1,j+1} \ e^{ i 2 (j+1) H }\ ,\label{def1}\\
&&C^{(2)}_{j,k}\ \a_1 \alpha_2\ \psi_{\frac{k}{2}-j-1;-\frac{k}{2}+j+2,-\frac{k}{2}+j+1}\
 \pi_{j;j+1,j+2}\
 e^{ i 2 (j+1) H_{\mathrm{L}}+
i (2(j+1)-k) H_{\mathrm{R}}}\ ,\label{def2}\\
&& C^{(3)}_{j,k}\ \alpha_2\a_1 \ \psi_{\frac{k}{2}-j-1;-\frac{k}{2}+j+1,-\frac{k}{2}+j+2}\
 \pi_{j;j+2,j+1}\
   e^{ i (2(j+1)-k) H_{\mathrm{L}}+i 2(j+1) H_{\mathrm{R}}}\ ,\label{def3}\\
&&C^{(4)}_{j,k}\ \alpha_2^2 \  \psi_{\frac{k}{2}-j-1;-\frac{k}{2}+j+1,-\frac{k}{2}+j+1}\
 \pi_{j;j+2,j+2}\
 e^{ i (2(j+1)-k) H}\label{def4}\ ,
\end{eqnarray}
where $H$ is a free boson defined as
\begin{equation}
H=\frac{1}{\sqrt{k}}\Big(\frac{P}{\sqrt{k-2}}+\frac{Q}{\sqrt{k+2}}\Big)\
. \label{dlkH}
\end{equation}
The $C^{(a)}_{j,k}$'s are numbers that depend on the relative
cocycles that we should include in principle when bosonizing the
fermions of the supersymmetric cosets. Their explicit form is not
necessary for our purposes, as we will be interested in the
large-$k$ limit in which we can infer easily that these numbers
for $a=1,2,4$ can be taken to be 1 and for $a=3$ to be $-1$. This
is essentially due to the fact that in the third term in \eqn{gd}
the order of the exponentials corresponding to bosonized fermions
for the compact and the non-compact cosets has been interchanged
compared to the third and fourth terms, whereas in \eqn{def3} the
order has been restored.

\no The four operators above are partners under ${\cal N}=1$
worldsheet supersymmetry since the deformations under
consideration should leave space--time supersymmetry (and hence
worldsheet supersymmetry as well) unbroken. In the semiclassical
limit $k\rightarrow \infty$ and for $j<<k$, where we can think of
them as deformations of the supersymmetric $\sigma$-model
Lagrangian with metric (\ref{sigmacirclecft}), the first one, Eq.
(\ref{def1}), would be a purely bosonic deformation. Indeed, the
contribution of the fermions, captured by the exponential
involving the bosonized field $H$, would be vanishing.
Accordingly, the operators (\ref{def2}) and (\ref{def3})
correspond to 2-fermion terms while (\ref{def4}) is a 4-fermion
interaction term. In the next section we will identify explicitly
the semiclassical limit of these operators with the corresponding
$\sigma$-model deformations. An exception to this picture would be
the case of $j=(k-2)/2$ which is of order $k$. This will be
treated separately and we will show that it corresponds also to a
nice geometrically interpreted deformation.

\subsection{Holographic dictionary: applications}

It is time now to employ the holographic dictionary presented above
in order to uncover how changing the configuration of the NS5-branes affects the
underlying conformal field theory description.

\subsubsection{Point to circle}

We start with a warm-up example taken from \cite{Aharony:2003vk}. Let us
distribute the NS5-branes symmetrically on a circle of radius $r_0$
on the $B$-plane so that  the eigenvalues $b_n$ take the form
\begin{equation}\label{vevcircle}
b_n=r_0\ e^{\frac{2\pi i}{k} n}\ .
\end{equation}
Since ${\rm tr}\langle B^l\rangle=0$ for $l<k$ we have
\begin{equation}\label{lambdajcircle}
\lambda_j=\mu \delta_{j,\frac{k-2}{2}}\ ,
\end{equation}
with $\mu \sim r_0^k$. The value of the spin $j=(k-2)/2$ is very special since
the multi-trace contributions, mentioned in footnote 3, are vanishing.
The deformed worldsheet theory is
\begin{equation}\label{Lpertcircle}
{\cal L}={\cal L}_0+\left( \mu \int d^2\theta e^{-\frac{1}{q}\Phi}
+{\rm c.c.}\right)\ ,
\end{equation}
which we recognize as the ${\cal N}=2$ Liouville theory. Combined with the
${\cal N}=2$ minimal model and upon orbifolding with $\mathbb{Z}_k$,
the theory thus obtained is equivalent to (\ref{cftcircle}), which indeed
describes the near-horizon geometry
of a circular configuration of NS5-branes \cite{Sfetsos:1998xd}.

\subsubsection{Circle to circle}\label{sec:step222}

It is interesting to consider the operator with $j=\frac{k-2}{2}$,
but in the non-singular  background  (\ref{cftcircle}). This
operator should correspond to changes of the radius of the
circular distribution of the NS5-branes. If we change the radius
from $r_0$ to $r_0+\delta$, we can either add to the original CHS
theory the $j=\frac{k-2}{2}$ operator with coefficients $\mu \sim
r_0^k$ and $\mu \sim  (r_0+\delta)^k$ respectively or, equivalently, deform the
$SL(2,\mathbb{R})/U(1) \times SU(2)/U(1)$ with the corresponding
operator bearing a coefficient $(r_0+\delta)^k-r_0^k \simeq k
r_0^{k-1} \delta$.

\no The operator we use in this case in \eqn{wdcircle} reads:
\begin{equation}\label{CFTpertcircle1}
V^{ su}_{0;0,0} V^{ sl}_{\frac{k}{2}-1;\frac{k}{2}, \frac{k}{2}}\
.
\end{equation}
Note that in this
case the spin $j$ is of the same order of magnitude as $k$, but as long as we stay within the
exact framework this does not present a problem. The correspondence with the semiclassical expressions
for the primaries will be done soon using the equivalence relation \eqn{euivv}.

\no
Since for this value of $j$ the coefficients $\alpha_1$
vanishes, the worldsheet Lagrangian changes by
\begin{equation}
\alpha_2^2 \psi_{0;0,0} \pi_{\frac{k}{2}-1;\frac{k}{2}+1,\frac{k}{2}+1} + {\rm c.c.} =
\alpha_2^2 \pi_{\frac{k}{2}-1;\frac{k}{2}+1,\frac{k}{2}+1} + {\rm c.c.}\ ,
\end{equation}
since $\psi_{0;0,0}$ is the identity field. It can be verified
using  (\ref{h25}) that this operator is marginal. Notice also
that this is a purely bosonic deformation, since the bosonic field
$H$ relating to the fermions doesn't appear. This is an explicit
example which indicates, among other things, that the r\^ole of
the various terms in (\ref{def1})-(\ref{def4}) for low spin when
$j\ll k$, as explained before the beginning of this subsection,
could be completely different for high spins having $j\sim
\frac{k}{2}$.

\no We would like now to find  a semiclassical expression for the
non-compact parafermionic primary $
\pi_{\frac{k}{2}-1;\frac{k}{2}+1,\frac{k}{2}+1}$. Notice that
(\ref{ncparapriOPE}) yields
\begin{equation}
 \pi (z) \pi_{\frac{k}{2}-1;\frac{k}{2},\frac{k}{2}} (w,\bar w)
\sim \frac{1}{(z-w)^{\frac{2}{k+2}}}
  \pi_{\frac{k}{2}-1;\frac{k}{2}+1,\frac{k}{2}+1} (w,\bar w)\ ,
\end{equation}
and similarly with $\pi(z)$ replaced by $\bar \pi(\bar z)$.
In the semiclassical limit $k\rightarrow \infty$ we can think of
$\pi_{\frac{k}{2}-1;\frac{k}{2}+1,\frac{k}{2}+1}$ as a
composite field $\pi \bar \pi \pi_{\frac{k}{2}-1;\frac{k}{2},\frac{k}{2}}$ and their
conformal dimensions match only to leading order in $1/k$, as it should be.
% as it should be for composite since the conformal weight of the latter is
%$1+\frac{2}{k+2}$ and agrees with that of $   \pi_{\frac{k}{2}-1;\frac{k}{2}+1,\frac{k}{2}+1}$
%for large $k$.
Now, the equivalence between non-compact parafermionic primaries can be
used to replace $  \pi_{\frac{k}{2}-1;\frac{k}{2},\frac{k}{2}}$ by $\pi_{0;-1,-1}$.
This is necessary as the semiclassical limit of the expression for the
primary  $  \pi_{\frac{k}{2}-1;\frac{k}{2},\frac{k}{2}}$ does not seem to be well-defined since
its spin is of order $k$.
The punch-line is that for large $k$ the leading deformation of the worldsheet Lagrangian
is through the operator
\begin{equation}
\alpha_2^2 \pi \bar \pi \pi_{0;-1,-1} + {\rm c.c.} \sim \alpha_2^2
\left(\frac{\partial\rho\bar \partial\rho}{\cosh^2\rho}-
\frac{\partial\omega\bar\partial\omega}{\sinh^2\rho}\right)\ ,
\label{ctc}
\end{equation}
where we have used the expressions \eqn{nonc1}, \eqn{nonc2} and
(\ref{prim1}) for the semiclassical $SL(2,\mathbb{R})/U(1)$
parafermions and primaries.

\no The above operator is a deformation of the $\sigma$ model
(\ref{sigmacirclecft}) and more precisely only of its
$SL(2,\mathbb{R})/U(1)$ part.  The $\sigma$ model of the latter is
defined in \eqn{companonco} and appending (\ref{ctc}) to it with
coefficient $\frac{1}{k} \alpha_2^2 k r_0^{k-1}\delta \sim
\epsilon k$ changes the above data to
\begin{equation}\label{sigdef}
\frac{ds^2}{k}=\left(1+ \frac{\epsilon}{\cosh^2 \rho}\right)
d\rho^2 + \left(\coth^2 \rho-\frac{\epsilon}{ \sinh^2 \rho}\right)
d\omega^2 ,\quad e^{-2 \Phi}= \sinh^2 \rho\ .
\end{equation}
Now, a coordinate redefinition $\rho \rightarrow \rho - \frac{\epsilon}{2}\tanh\rho$
brings the deformed metric back to its original form but changes the profile
of the dilaton as
\begin{equation}\label{dilshtru}
e^{-2 \Phi}= (1-\epsilon) \sinh^2 \rho\ .
\end{equation}
In the T-dual theory, i.e.~the cigar, the dilaton takes the form
\begin{equation}\label{dilshcig}
e^{-2 \Phi}= (1-\epsilon) \cosh^2 \rho\
\end{equation}
and its value at $\rho=0$, i.e. at the tip of the cigar,
 is related to the radius of the circle on which we
put the NS5-branes \cite{Giveon:1999zm}.
Here we see explicitly that changing this radius has
the expected effect on the value of the dilaton at the tip.

\subsubsection{Circle to ellipsis}

Let us now consider a deformation of the circular configuration of
NS5-branes to an ellipsis.
%The background for the distribution of NS5-branes on an ellipsis was constructed
%in \cite{Marios Petropoulos:2005wu} and for small deviations from the circular distribution it was
%shown that the perturbation can be written as
If the latter has radii $r_1=r_0(1+\epsilon)$ and
$r_2=r_0(1-\epsilon)$ the positions of the NS5-branes on the
$x^8-x^9$ plane are parameterized by $\langle \Phi^8_n\rangle =r_1
\cos(\phi_n), \left\langle \Phi^9_n\right\rangle =r_2
\sin(\phi_n)$ with $\phi_n=\frac{2\pi}{k}n, n=1,\ldots,k$. Since
$\langle B\rangle=\left\langle\Phi^8\right\rangle+i
\left\langle\Phi^9\right\rangle$ we have
\begin{equation}\label{vevellipsis}
b_n= r_0\left( e^{\frac{2\pi i}{k}n}+{\epsilon }  e^{-\frac{2\pi
i}{k}n}\right)\ .
\end{equation}
We can think of $\epsilon$ as controlling the deformation of a
circle of radius $r_0$ to an ellipsis. The couplings $\lambda_j$
are proportional to
\begin{equation}\label{Bmellipis}
\left\langle {\rm tr} (B)^m\right\rangle = k r^m\sum_{l=0}^m {m
\choose l}  \epsilon^l
 (\delta_{2l-m,k}+\delta_{2l,m}+\delta_{2l-m,-k})\ ,
\end{equation}
where $m=2j+2=2,3,\ldots,k$. Operators with a given $j \in
\mathbb{N}$ result in corrections of order $\epsilon^{j+1}$. The
leading deformation, which is of order $\epsilon$, appears for
$m=2 $ and therefore $j=0$. The corresponding operator we use in
\eqn{wdcircle} is
\begin{equation}\label{ellipsisdef}
V^{ su}_{\frac{k}{2}-1;-\frac{k}{2}+1,-\frac{k}{2}+1} V^{
sl}_{0;1,1}
\end{equation}
and we can read from  (\ref{def1})--(\ref{def4}) the associated
Lagrangian deformations:
\begin{equation}
\label{defel3}
\begin{array}{rcl}
&&\displaystyle{C^{(1)}_{0,k}\ \alpha_1^2\
\psi_{\frac{k}{2}-1;-\frac{k}{2}+2,-\frac{k}{2}+2}\
 \pi_{0;1,1} \ e^{  2 i   H }}\ , \crbig
&&\displaystyle{C^{(2)}_{0,k}\ \a_1\alpha_2 \
\psi_{\frac{k}{2}-1;-\frac{k}{2}+2,-\frac{k}{2}+1}\
 \pi_{0;1,2}\
 e^{ 2 i  H_{\mathrm{L}}+
i (2-k) H_{\mathrm{R}}}}\ , \crbig &&\displaystyle{C^{(3)}_{0,k}\
\alpha_2\a_1 \ \psi_{\frac{k}{2}-1;-\frac{k}{2}+1,-\frac{k}{2}+2}\
 \pi_{0;2,1}\
   e^{ i (2-k) H_{\mathrm{L}}+ 2 i H_{\mathrm{R}}}}\ , \crbig
&&\displaystyle{C^{(4)}_{0,k}\ \alpha_2^2 \
\psi_{\frac{k}{2}-1;-\frac{k}{2}+1,-\frac{k}{2}+1}\
 \pi_{0;2,2}\
 e^{ i (2-k) H}}\ .
\end{array}
\end{equation}
Recall also that we have to add to the operators above their complex conjugates.

 \no
We would like now to understand the semiclassical limit of these
operators. We begin with the exponentials and in the large-$k$
limit we have the following correspondences

\begin{equation}
\label{cdjk1}
\begin{array}{rclrcl}
e^{2 i H}&\to &1\ , &\quad h&=&\bar h \simeq 0\ , \crbig e^{ 2 i
H_{\mathrm{L}}+ i (2-k) H_{\mathrm{R}}}&\to & e^{ -i
(P_{\mathrm{R}}+Q_{\mathrm{R}})}\ , &\quad h&\simeq &0\ ,\quad
\bar h \simeq \displaystyle{1 -{4\ov k}}\ , \crbig
  e^{  i (2-k) H_{\mathrm{L}}+ i 2 H_{\mathrm{R}}} &\to & e^{ -i (P_{\mathrm{L}}+Q_{\mathrm{L}})}\ , &\quad h&\simeq & \displaystyle{1 -{4\ov k}}\
  ,\quad \bar h
\simeq 0\ , \crbig e^{i(2-k) H} &\to & e^{ -i (P+Q)}\ , &\quad
h&=&\bar h \simeq \displaystyle{1 -{4\ov k} }\ ,
\end{array}
\end{equation}
where we have indicated the semiclassical expressions explicitly
in terms of the original bosons $P$ and $Q$,
along with the conformal dimension up to
order ${\cal O}(1/k)$ for which we are interested in.

 \no
 For the non-compact parafermionic primaries the semiclassical
 expressions can be read from (\ref{prim1}) and (\ref{prim2}).
 For the compact parafermionic primaries similar formulas would be ill-defined
 semiclassicaly (an analogous  situation was encountered for the non-compact primaries
 in the previous example) and we should find them indirectly. For instance
 the dimension of $  \psi_{\frac{k}{2}-1;-\frac{k}{2}+1,-\frac{k}{2}+1}$ is $0$
 and indeed the parafermionic equivalence (\ref{cpeq}) relates this operator to the identity
 $\psi_{0;0,0}$. Furthermore, for  $\psi_{\frac{k}{2}-1;-\frac{k}{2}+2,-\frac{k}{2}+1}$ the
dimension of the left part is $h=1-1/(k-2)$, i.e.~the same as the one for the
compact parafermions and moreover, using the OPE (\ref{cparapriOPE}) (for $j=k/2-1$,
$m=-k/2+1$ and remembering first to shift in that expression $k\to k-2$)
 we conclude that it can be
identified with that. Hence, we have the exact identifications
 \begin{equation}
\psi_{\frac{k}{2}-1;-\frac{k}{2}+2,-\frac{k}{2}+2} =\psi \bar \psi\ , \quad
 \psi_{\frac{k}{2}-1;-\frac{k}{2}+2,-\frac{k}{2}+1} = \psi\ ,  \quad
 \psi_{\frac{k}{2}-1;-\frac{k}{2}+1,-\frac{k}{2}+2} = \bar \psi\ .
 \end{equation}

 \no
 We can now assemble everything using the expressions (\ref{classcp}) and
 (\ref{classcp1}) for the parafermion
 fields and (\ref{prim1}) and (\ref{prim2}) for the non-compact primaries.
 The deforming operators in the semiclassical
 limit read:
 \begin{equation}
\label{petrre}
\begin{array}{rcl}
 && \displaystyle{k\, {1\ov g_{--}^2}\, \psi\bar \psi}\ , \crbig
 && \displaystyle{- {2 i\ov g_{--}^3} \left( g_{-+}\ \psi\,
e^{-i(P_{\mathrm{R}}+Q_{\mathrm{R}})}  +
 g_{+-}\, \bar\psi\ e^{-i(P_{\mathrm{L}}+Q_{\mathrm{L}})} \right)} \ ,\crbig
 && \displaystyle{{2\ov k}\ {2 g_{+-}g_{-+}-1\ov g_{--}^4}\,  e^{-i(P+Q)}}\ ,
\end{array}
\end{equation}
where, compared with \eqn{cdjk1},
 we have replaced the constant factors $a_{1,2}$ by their leading $1/k$ behavior and in addition
we have set the co-cycle dependent factors $C^{(a)}_{0,k}$ to
their semiclassical values as described below \eqn{dlkH}. It is
understood that $\psi,\bar \psi$ and the various group elements
are represented  in terms of the target-space variables as in Sec.
\ref{sec:step11}. In the next section we will identify them with
the ${\cal N}=1$ $\sigma$ model deformations induced by changing
the circular distribution of NS5-branes to an elliptical one. In
particular, the first line above will be related to the
deformation of the bosonic part, thus recovering the result in
\cite{Marios Petropoulos:2005wu}, the second line will correspond
to the deformation of the fermion bilinears and the third line to
the deformation of the quartic, in the fermions, term.

\subsubsection{General circle deformations}

We consider now general deformations of a circular distribution of NS5-branes.
Their positions in the $B$-plane are encoded by
\begin{equation}\label{genvev}
b_n=r_0 \left(e^{\frac{2 \pi i n}{k}} +  \epsilon_n\right)\ ,\quad
n=1,\ldots,k\ ,\qq \sum_{n=0}^{k} \epsilon_n =0\ .
\end{equation}
Here $\epsilon_n << 1 $ are complex numbers indicating the shift
of the position of the $n$-th NS5-brane relative to the original circle.
To leading order in $\epsilon_n$ all operators in (\ref{circlecftchiral})
for which the sum
\begin{equation}\label{gdpar}
\sum_{n=0}^{k} \epsilon_n\ e^{\frac{2 \pi i n (2j+1)}{k}}
\end{equation}
is non-zero, contribute.
%For instance, the operator with $j=0$ contributes with parameter
% \begin{equation}\label{genex}
% \sum_{n=0}^{k-1} \epsilon_n e^{\frac{2 \pi i n}{k}}
 %\end{equation}
%and this is indeed non-vanishing for the case of the elliptic deformation
%considered previously.

\no
It is natural to proceed by looking for a deformation of the circle distribution
so that only a single operator is turned on. In other words, we would like a set
of numbers $\epsilon_n$ with the property that their sum is zero
and all sums of the form (\ref{gdpar}) are zero except for one given value of $j$.
Obviously such a set exists and it is given by $\epsilon_n=
\epsilon\ e^{- \frac{2 \pi i n (2j+1)}{k}}$. For instance, for $j=0$
we retrieve the deformation to the ellipsis studied previously,
while for $j=(k-2)/2$ we see that, as expected, we change the
radius $ r_0 \rightarrow r_0(1+\epsilon)$.
The perturbed positions of the NS5 branes are
\begin{equation}\label{genvev1}
b_n=r_0 \left(e^{\frac{2 \pi i n}{k}} + \epsilon\ e^{- \frac{2 \pi i n
(2j+1)}{k}} \right) \ , \qq n=1,\ldots,k\ .
\end{equation}

\no An important feature of the distribution (\ref{genvev1}) is
that the NS5-branes are not uniformly distributed. In order to
understand that, we should consider the continuum limit of
(\ref{genvev1}), which will be also important for the
considerations of the next section. In this limit $k\rightarrow
\infty$ and the discrete angles $\phi_n=\frac{2 \pi n}{k}$ are
replaced by a continuous angle $\phi\in[0,2\pi]$. The discrete distribution
(\ref{genvev1}) approaches
\begin{equation}
b(\phi)=r_0 e^{i \phi} \left(1+\epsilon\ e^{-2 i (j+1)\phi}\right)
\label{jk99}
\end{equation}
and to leading order we can think of that as deforming the radius
$r_0$ of the original circular distribution as $r_0 \rightarrow
r_0(1+ \epsilon \cos 2(j+1)\psi)$ and also changing the angle as
$\phi= \psi +\e \sin 2(j+1)\psi$. Hence the original uniform
angular distribution with measure $d\psi$ is replaced by
\begin{equation}
d\psi\ \l(\psi)\ ,\qq \l(\psi)= 1 +  2\e\ (j+1) \cos 2(j+1)\psi\ .
\label{meas}
\end{equation}

\no We would like now to understand the semiclassical limit of the
general operators \eqn{def1}-\eqn{def4}. As in the case of the
elliptical deformation we begin with the exponentials and in the
large-$k$ limit we have the following correspondences
\begin{equation}
\label{dims}
\begin{array}{rclrcl}
e^{2 i (j+1) H}&\to &1\ , &\quad h&=&\bar h \simeq 0\ , \crbig
e^{ 2 i (j+1) H_{\mathrm{L}}+ i [2(j+1)-k] H_{\mathrm{R}}}&\to & e^{ -i
(P_{\mathrm{R}}+Q_{\mathrm{R}})}\ , &\quad h&\simeq &0\ ,\quad
\bar h \simeq \displaystyle{1 -{4(j+1)\ov k}}\ , \crbig
  e^{ i [ 2(j+1)-k] H_{\mathrm{L}}+  2 i (j+1) H_{\mathrm{R}}} &\to & e^{ -i (P_{\mathrm{L}}+Q_{\mathrm{L}})}\ , &\quad h&\simeq & \displaystyle{1 -{4(j+1)\ov k}}\
  ,\quad \bar h
\simeq 0\ , \crbig e^{i [ 2(j+1)-k]H} &\to & e^{ -i (P+Q)}\ , &\quad
h&=&\bar h \simeq \displaystyle{1 -{4(j+1)\ov k} }\ ,
\end{array}
\end{equation}
where we have indicated the semiclassical expressions explicitly
in terms of the original bosons $P$ and $Q$,
along with the conformal dimension up to
order ${\cal O}(1/k)$ for which we are interested in.

\no
Considering the compact primaries,
as in the previous example of the elliptical deformation, we will find them indirectly, but
extra care is needed as some of them arise as composite operators.
The parafermionic
equivalence (\ref{cpeq}) implies that
$ \psi_{\frac{k}{2}-j-1;-\frac{k}{2}+j+1,-\frac{k}{2}+j+1}$ is related to
the operator $\psi_{j;j,j}$.
Furthermore, the primary $\psi_{\frac{k}{2}-j-1;-\frac{k}{2}+j+2,-\frac{k}{2}+j+1}$ arises as
a composite operator of the
compact parafermion $\psi$ with $\psi_{j;j,j}$. This can be seen by using the OPE
\be
\psi(z) \psi_{{k\ov 2}-j-1;-{k\ov 2}+j+1,-{k\ov 2}+j+1}(w,\bar w) \sim {k-2(j+1)\ov \sqrt{k-2}}
 (z-w)^{-2j\ov k-2}\
\psi_{{k\ov 2}-j-1;-{k\ov 2}+j+2,-{k\ov 2}+j+1}(w,\bar w)\ ,
\ee
arising from (\ref{cparapriOPE})
(remembering to shift $k\to k-2$) and the parafermionic equivalence.
Hence, we have the exact identifications
\begin{equation}
\begin{array}{rclrcl}
\psi_{\frac{k}{2}-j-1;-\frac{k}{2}+j+2,-\frac{k}{2}+j+2} &=&\psi_{j;j,j}\psi \bar \psi\ ,&\qq h&=&\bar h\simeq \displaystyle{
1-{j+1\ov k}}\ ,
\crbig
 \psi_{\frac{k}{2}-j-1;-\frac{k}{2}+j+2,-\frac{k}{2}+j+1}& =& \psi_{j;j,j}\psi \ ,
&\qq h&\simeq& \displaystyle{1-{j+1\ov k}}\ ,\ \bar h= {j\ov k}\ , \crbig
\psi_{\frac{k}{2}-j-1;-\frac{k}{2}+j+1,-\frac{k}{2}+j+2}& =&
\psi_{j;j,j}\bar \psi\ ,\qq &\qq h&=&\displaystyle{{j\ov k}} \ ,\ \bar h\simeq
\displaystyle{1-{j+1\ov k}} \ ,
\end{array}
\end{equation}
where we have indicated the
conformal dimensions of the different terms. We emphasize that, in
the product of two operators of the compact coset the dimension is
not just the sum of the dimensions. One actually should think of
them as one composite operator defined through the OPE
\eqn{cparapriOPE}. The above identifications are correct as merely
a simple product, when we will later perform the comparison with
the $\cN=1$ $\s$ model.

 \no
 We can now assemble everything using the expressions (\ref{classcp}) and
 (\ref{classcp1}) for the parafermion
 fields and (\ref{prim1}) and (\ref{prim2}) for the non-compact primaries.
 The deforming operators in the semiclassical
 limit read:
 \begin{equation}
\label{hologenop}
\begin{array}{rcl}
 && \displaystyle{k\, {\tilde g_{+-}^{2j}\ov g_{--}^{2(j+1)}}\, \psi\bar \psi}\ , \crbig
 && \displaystyle{- 2(j+1)\, {i\tilde g_{+-}^{2j} \ov g_{--}^{2j+3}} \left( g_{-+}\,
\psi\ e^{-i(P_{\mathrm{R}}+Q_{\mathrm{R}})}  +
 g_{+-}\, \bar\psi\ e^{-i(P_{\mathrm{L}}+Q_{\mathrm{L}})} \right)} \ ,\crbig
 && \displaystyle{{2(j+1)\ov k}\, \tilde g_{+-}^{2j} {2(j+1) g_{+-}g_{-+}-1\ov g_{--}^{2(j+2)}}\,
 e^{-i(P+Q)}}\ ,
\end{array}
\end{equation}
to which we should add their complex conjugates.

\section{\boldmath Deformations of NS5-branes: $\sigma$ model
approach and comparison\unboldmath}\label{sec:step3}

In this section we perform a complete comparison between a general
infinitesimal deformation of a circular distribution of NS5-branes
on the same plane, as it results from two different approaches.
Namely, either from the corresponding $\cN=1$ supersymmetric $\s$
model viewpoint or from the exact CFT approach of the previous
section  as an exactly marginal perturbation based on the
holographic conjecture. First we consider the simpler case of the
circular distribution deformed into a small ellipsis and complete
the analysis of \cite{Marios Petropoulos:2005wu} which was
performed only for the bosonic part of the supersymmetric action.
Subsequently, we consider the general planar deformation. The two
approaches are in complete agreement in their semiclassical range
of common validity, including numerical coefficients.

\subsection{Neveu--Schwarz five-branes on an ellipsis}

Let us consider a system of $k$  NS5-branes distributed on the circumference of an ellipsis
with axes $\ell_1$ and $\ell_2$ according to the density
\begin{equation}
\rho({\bf x})=\frac{1}{\pi\ell_1\ell_2}
\delta\left(1-\frac{\left(x^8\right)^2}{\ell_1^2}-\frac{\left(x^9\right)^2}{\ell_2^2}\right)
\delta\left(x^6\right)\delta\left(x^7\right) \ . \label{elde}
\end{equation}
The supergravity solution corresponding to (\ref{elde}) along with
its T-dual was constructed and studied in detail in \cite{Marios
Petropoulos:2005wu}. Here we are interested in the infinitesimal
deformation of a circular distribution with $\ell_1=\ell_2$, which
admits the exact CFT description (\ref{cftcircle}), to an ellipsis
with $\ell_1=r_0(1+\epsilon), \ell_2= r_0(1-\epsilon)$. Given a
density distribution one can construct the corresponding
supergravity solution from \eqn{met1}-\eqn{dill}. In our case it
is convenient to parameterize the Cartesian coordinates in a way
appropriate to the density \eqn{elde} as \cite{Marios Petropoulos:2005wu}
\begin{equation}
{\bf x} = \ell_1 \left( \sinh\rho \cos \theta \cos \tau,
 \sinh\rho \cos \theta \sin \tau,
  \cosh\r \sin \theta \cos \psi,
 \sqrt{\sinh^2\r + \frac{\ell_2^2}{\ell_1^2}}\ \sin \theta \sin \psi\right)  \ .\label{x4ell}
\end{equation}
In the limit $\ell_1=\ell_2=r_0$ one obtains the supergravity
solution representing a uniform continuous NS5-brane distribution
\cite{Sfetsos:1998xd} which semiclassicaly is described, after a
proper T-duality, by the $\sigma$ model (\ref{sigmacirclecft}).

\no The leading order correction to  (\ref{sigmacirclecft}),
corresponding to an infinitesimal deformation of the circle
towards an ellipsis and after the proper T-duality and a
reparameterization, reads  \cite{Marios Petropoulos:2005wu} (see
Eq. (38) in that Ref.):
\begin{equation}
 ds^{(1)2}
= \frac{2\epsilon k}{\cosh^2 \rho} \big[\cos
2(\omega-\varphi)\left(d\th^2-\tan^2\th d\varphi^2\right)
 + 2\sin 2(\omega-\varphi) \tan\th d\varphi d\th \big] \ ,
\label{defmetel}
\end{equation}
where we note the relation of the angular variables as compared to
\eqn{x4ell} is  $\varphi=\tau$ and $\om=\tau + \psi$. The full
metric is the sum of the metric in  (\ref{sigmacirclecft}) and of that in
(\ref{defmetel}), whereas the antisymmetric tensor remains
vanishing. It will be convenient to work with a coframe $\{e^{\hat
\imath}\}$ such that the full metric takes the form
\begin{equation}\label{metric1}
ds^{2} = \eta_{\hat \imath\hat \jmath} e^{\hat \imath} e^{\hat
\jmath}\ ,
\end{equation}
with the hatted indices taking values $\hat \imath=\hat
1,\ldots,\hat 4$ and where
\begin{equation}\label{tsmetric}
\eta_{\hat \imath\hat \jmath}=
\begin{pmatrix}
0 & {1 \ov 2} & 0 & 0 \\
\frac{1}{2} & 0 & 0 & 0 \\
0 & 0 & 0& \frac{1}{2} \\ 0 & 0  & \frac{1}{2}& 0
\end{pmatrix}\ .
\end{equation}
The relevant coframe, associated spin connections and curvature
tensors can be retrieved from App. \ref{geodata} by setting $n=2$
as it can be easily seen by comparing (\ref{defmetel}) with
(\ref{D1}).

\no The ${\cal N}=1$ $\sigma$ model using tangent-space objects is
\cite{zumino, gaume,dz,howe}
\begin{eqnarray}
S  =  \int  d^2\sigma &&\bigg(\eta_{\hat \imath \hat \jmath}
e_-^{\hat \imath} e_+^{\hat \jmath}+ i  \eta_{\hat \imath\hat
\jmath}  \Psi^{\hat \imath}_+ \Big(\partial_- \Psi^{\hat \jmath}_+
+ \omega^{\hat \jmath}_{-{\hat k}} \Psi^{\hat k}_+\Big)
\nonumber\\
&&+ i  \eta_{\hat \imath\hat \jmath}  \Psi^{\hat \imath}_-
\Big(\partial_+ \Psi^{\hat \jmath}_- + \omega^{\hat
\jmath}_{+{\hat k}} \Psi^{\hat k}_-\Big)+ \frac{1}{2} R_{\hat
\imath \hat \jmath\hat k\hat l} \Psi^{\hat \imath}_+ \Psi^{\hat
\jmath}_+ \Psi^{\hat k}_- \Psi^{\hat l}_- \bigg)\ ,
 \label{feraction}
\end{eqnarray}
where $e^{\hat \imath}_{\pm}$ are the ``chiral coframes", i.e.~the
coframes with their exterior differentials replaced by the
appropriate chiral derivative, and similarly $ \omega^{\hat
\jmath}_{\pm \hat k}$ is the corresponding connection $
\omega^{\hat \jmath}_{\pm \hat k} = \omega^{\hat
\jmath}_{\hphantom{j}\hat \imath\hat k}e^{\hat \imath}_\pm$.
Notice that the connection is the usual Levi--Civita connection
since we have vanishing NS--NS three-form flux.

\no
 The fermions in (\ref{feraction}) are coupled to the scalar sector of the theory.
 Hence, we would like to perform a field
redefinition that renders them free, in the undeformed background,
as in the CFT formulation in \cite{Aharony:2003vk}. Therefore, we
write the ${\cal N}=1$ $\sigma$ model (\ref{feraction}) using the
unperturbed (i.e.~$\epsilon=0$) connections from
(\ref{connection}) and curvature tensors from (\ref{curv2form}) as
\begin{eqnarray}\label{sigmacircle}
S_0= \int d^2\sigma &&\bigg(\Big(k\left(\partial_+ \th
\partial_-\th + \tan^2 \th
\partial_+\varphi \partial_-\varphi\right)  +  {i \over 2} \left(\Psi^{\hat{3}}_
+ \partial_- \Psi^{\hat{4}}_+ + \Psi^{\hat{4}}_+ \partial_-
\Psi^{\hat{3}}_+\right) +  i\Psi^{\hat{3}}_+ \Psi^{\hat{4}}_+ \omega^{\hat{4}}_{-\hat{4}} \nonumber \\
&&+{i\over 2} \left(\Psi^{\hat{3}}_- \partial_+ \Psi^{\hat{4}}_- +
\Psi^{\hat{4}}_+
\partial_- \Psi^{\hat{3}}_+\right)   +  i \Psi^{\hat{3}}_-
\Psi^{\hat{4}}_- \omega^{\hat{4}}_{+\hat{4}} + 2
R_{\hat{3}\hat{4}\hat{3}\hat{4}} \Psi^{\hat{3}}_-
\Psi^{\hat{4}}_-\Psi^{\hat{3}}_+ \Psi^{\hat{4}}_+ \Big)
 \nonumber\\
&&+\Big(k\left(\partial_+ \rho \partial_-\rho + \coth^2 \rho
\partial_+\omega \partial_-\omega\right)
 +  {i \over 2} \left(\Psi^{\hat{1}}_+ \partial_- \Psi^{\hat{2}}_+ +
\Psi^{\hat{2}}_+ \partial_- \Psi^{\hat{1}}_+\right) + i
\Psi^{\hat{1}}_+ \Psi^{\hat{2}}_+ \omega^{\hat{2}}_{-\hat{2}}
\nonumber\\
&&+{i\over 2} \left(\Psi^{\hat{1}}_- \partial_+ \Psi^{\hat{2}}_- +
\Psi^{\hat{2}}_+
\partial_- \Psi^{\hat{1}}_+\right)  +
i \Psi^{\hat{1}}_- \Psi^{\hat{2}}_- \omega^{\hat{2}}_{+\hat{2}} +
2 R_{\hat{1}\hat{2}\hat{1}\hat{2}} \Psi^{\hat{1}}_-
\Psi^{\hat{2}}_-\Psi^{\hat{1}}_+ \Psi^{\hat{2}}_+\Big) \bigg)\
.\label{xeroo}
\end{eqnarray}
The fermions $\Psi_\pm^{\hat 1}$ and $\Psi_\pm^{\hat
2}=\big(\Psi_\pm^{\hat 1}\big)^*$ belong to the non-compact coset
while $\Psi_\pm^{\hat 3}$ and $\Psi_\pm^{\hat
4}=\big(\Psi_\pm^{\hat 3}\big)^*$ belong to the compact one.

\no
Now, the field redefinitions
\begin{equation}\label{FR1}
 \partial_\pm \varphi \to \partial_\pm\varphi + {1\ov k} \Psi^3_\pm\Psi^4_\pm\ ,\qq
 \partial_\pm \omega \to \partial_\pm\omega -{1\ov k} \Psi^1_\pm\Psi^2_\pm\ ,
\end{equation}
when substituted into \eqn{xeroo}, lead classically to the
following $\sigma$ model
\begin{eqnarray}\label{sigma3}
S = \int  d^2 \sigma &&\bigg( k \left(\partial_+ \th \partial_-\th
+ \tan^2 \th
\partial_+\phi
\partial_-\phi\right)  +  {i\over 2} \left(\Psi^{\hat{3}}_+ \partial_- \Psi^{\hat{4}}_+ + \Psi^{\hat{4}}_+ \partial_-
\Psi^{\hat{3}}_+\right)
\nonumber\\
&& +{i\over 2} \left(\Psi^{\hat{3}}_- \partial_+ \Psi^{\hat{4}}_-
+ \Psi^{\hat{4}}_- \partial_+ \Psi^{\hat{3}}_-\right)  +  {1\ov
k}\Psi^{\hat{3}}_- \Psi^{\hat{4}}_-\Psi^{\hat{3}}_+
\Psi^{\hat{4}}_+
 \nonumber\\
&& + k \left(\partial_+ \rho \partial_-\rho + \coth^2 \rho
\partial_+\omega
\partial_-\omega\right)  +  {i \over 2} \left(\Psi^{\hat{1}}_+ \partial_- \Psi^{\hat{2}}_+
+ \Psi^{\hat{2}}_+ \partial_- \Psi^{\hat{2}}_+\right)\nonumber
\\
&& +{i\over 2} \left(\Psi^{\hat{1}}_- \partial_+ \Psi^{\hat{2}}_-
+ \Psi^{\hat{2}}_- \partial_+ \Psi^{\hat{1}}_-\right)  - {1\ov
k}\Psi^{\hat{1}}_- \Psi^{\hat{2}}_-\Psi^{\hat{1}}_+
\Psi^{\hat{2}}_+ \bigg)\ .
\end{eqnarray}
The fermions have been decoupled from the scalars and are subject only to
a Thirring-type interaction \cite{Bagger:1986cd, Kounnas:1993ix}. Therefore, they can be described through a
compact boson of radius  $R$ given by
\begin{equation}\label{thirring}
R-{1\over R}+2h=0\ ,
\end{equation}
where $h$ is the coupling constant of the 4-fermi interaction term. In our case
the free boson radii for the two cosets turn out to be
\begin{equation}
\begin{array}{crl}
\label{raddi} &&\displaystyle{R_{\nicefrac{SU(2)}{U(1)}}}=
\displaystyle{ -{1\ov k} + \sqrt{1+{1\over k^2}} \simeq 1 - {1\ov
k}\simeq \sqrt{\frac{k-2}{k}}}= \displaystyle{R^{\rm
CFT}_{\nicefrac{SU(2)}{U(1)}}}\ ,\crbig
&&\displaystyle{R_{\nicefrac{SL(2,\mathbb{R})}{U(1)}}}=
\displaystyle{{1\ov k} +
 \sqrt{1+{1\over k^2}} \simeq 1 + {1\ov k}\simeq
\sqrt{\frac{k+2}{2}}}= \displaystyle{R^{\rm
CFT}_{\nicefrac{SL(2,\mathbb{R})}{U(1)}}} \ .
\end{array}
\end{equation}
The CFT radii at the right-hand side are those of the bosons $P$
and $Q$ that bosonize the free fermions of each supersymmetric
coset and are consistent with the way these bosons enter into the exponentials in
\eqn{SUGpm} and \eqn{SLGpm}.
The match here is only to leading order in $1/k$  since the
$\sigma$ model captures only the semiclassical features of the
exact CFT and furthermore we have applied the field redefinitions
classically ignoring any Jacobians from the transformations. These
fermions correspond to the bosonized ones of the CFT side in terms
of the fields $P$ and $Q$ defined in Sec. \ref{sec:step12} as
follows
\begin{equation}
\label{bosone}
\begin{array}{crlcrlcrlcrl}
e^{i\phi_2} e^{-i Q_{\mathrm{L}}} &=&  \Psi^{\hat 1}_+\ ,&\quad
e^{-i\phi_2}e^{i Q_{\mathrm{L}}} &=& \Psi^{\hat 2}_+\ ,&\quad
e^{i\phi_1} e^{-i P_{\mathrm{L}}} &=&  \Psi^{\hat 3}_+\ ,& \quad
e^{-i\phi_1} e^{i P_{\mathrm{L}}} &=& \ \Psi^{\hat 4}_+\ ,\crbig
e^{-i\phi_2} e^{-i Q_{\mathrm{R}}} &=&  \Psi^{\hat 1}_-\ ,&\quad
e^{i\phi_2}e^{i Q_{\mathrm{R}}} &=& \Psi^{\hat 2}_-\ ,&\quad
e^{-i\phi_1} e^{-i P_{\mathrm{R}}} &=&  \Psi^{\hat 3}_-\ , &\quad
e^{i\phi_1} e^{i P_{\mathrm{R}}} &=& \ \Psi^{\hat 4}_-\ .
\end{array}
\end{equation}

\no The introduction  of the non-local phases is necessary in
order to ensure gauge invariance of the fermions that have their
origin in the ungauged theory. This is the same reason for the
similar non-local phases appearing in the definitions of the
classical parafermions. Since the fermions anticommute we should
have included a co-cycle factor in the above definitions. This is
not necessary as long as, in the comparison with the
supersymmetric $\s$-model results of the deformation below, we
keep the fermions of the compact coset $\Psi_\pm^{3,4}$ to the
left of the non-compact counterparts $\Psi_\pm^{1,2}$. In this
manner the effect of the non-trivial co-cycles in the
semiclassical limit we are interested in is properly taken into
account by the fact that in \eqn{defel3} we have
 $C^{(a)}_{0,k}=1$ for $a=1,2,4$ and $C^{(3)}_{0,k}=-1$ in that limit.

\no The idea now is to write the complete $\sigma$ model for the
deformed background, perform the field redefinitions (\ref{FR1})
to render the fermions free in the undeformed $\sigma$ model, and
consider the leading correction. The discussion will be more
transparent if we consider separately the purely bosonic piece,
the 2-fermi term, and the 4-fermi term. The purely bosonic piece
is
\begin{equation}
S_{\rm bose} = \int d^2\sigma \frac{1}{2} \left(e^{\hat 1}_+
e^{\hat 2}_- + e^{\hat 1}_-  e^{\hat 2}_+ + e^{\hat 3}_+ e^{\hat
4}_- +  e^{\hat 3}_- e^{\hat 4}_+\right)\label{bosel}
\end{equation}
and performing the redefinitions (\ref{FR1}) on this term does not
generate any extra bosonic terms. The order $\epsilon$ correction
in (\ref{bosel}) can be written very compactly if we notice that
the chiral coframes  $e^{\hat \imath}_{\pm}$ with $\hat
\imath=3,4$, which are the only ones contributing to the
correction, can be expressed in terms of the parafermion fields
(\ref{classcp}) and (\ref{classcp1})
as\footnote{Left-/right-moving fields labeled by $+/-$ correspond
to holomorphic/antiholomorphic objects.}
\begin{equation}
\begin{array}{crlcrl}
e^{\hat 3}_+ &=& e^{i \phi_1} \psi + \epsilon  \displaystyle{\frac{e^{-2i\omega}}{\cosh^2\rho}}
e^{-i \phi_1} \psi^\dagger\ ,\quad
e^{\hat 3}_- &=& e^{-i \phi_1} \bar \psi + \epsilon \displaystyle{\frac{e^{-2i\omega}}{\cosh^2\rho}
e^{i \phi_1}} \bar \psi^\dagger\ ,
\crbig
e^{\hat 4}_+ &=& e^{-i \phi_1} \psi^\dagger + \epsilon  \displaystyle{\frac{e^{2i\omega}}{\cosh^2\rho}}
e^{i \phi_1} \psi\ , \quad
e^{\hat 4}_- &=& e^{i \phi_1} \bar  \psi^\dagger +
 \epsilon  \displaystyle{\frac{e^{2i\omega}}{\cosh^2\rho}}
e^{-i \phi_1}\bar  \psi\ .
\end{array}
\end{equation}
Consequently, the order $\epsilon$ bosonic correction reads:
\begin{equation}
\delta S_{\rm bose} = \epsilon \int d^2\sigma
\frac{k}{\cosh^2\rho} \left(e^{2 i\omega}\psi \bar \psi + e^{-2 i
\omega} \psi^\dagger \bar \psi^\dagger\right)\ . \label{boscorel}
\end{equation}
Taking into account \eqn{ginva} and \eqn{bosone} we see that this is indeed the first line of the perturbation
\eqn{petrre} based on CFT considerations, with infinitesimal parameter $\e$. Of course it also
reproduces the change in the metric \eqn{defmetel} above, due to the deformation.
This matching of the bosonic
part of the action corresponding to the deformation  was shown in
\cite{Marios Petropoulos:2005wu}. We move on now to the fermionic pieces of the action.

 \no
The original 2-fermi terms are\footnote{
Notice that we have used symmetry properties like
$ \omega^{\hat{1}}_{-\hat{1}} = - \omega^{\hat{2}}_{-\hat{2}}$ etc.~to reduce the number of terms. }
 \begin{eqnarray}
S_{2-{\rm fermi}}=i \int  d^2\sigma  &&\bigg(
 \left(\Psi^{\hat{1}}_+ \Psi^{\hat{2}}_+ \omega^{\hat{2}}_{-\hat{2}} +
 \Psi^{\hat{1}}_- \Psi^{\hat{2}}_- \omega^{\hat{2}}_{+\hat{2}}\right) +
 \left(\Psi^{\hat{3}}_+ \Psi^{\hat{4}}_+ \omega^{\hat{4}}_{-\hat{4}} +
 \Psi^{\hat{3}}_- \Psi^{\hat{4}}_- \omega^{\hat{4}}_{+\hat{4}}\right) + \nonumber\\
&&+ \left(\Psi^{\hat{2}}_+ \Psi^{\hat{4}}_+
\omega^{\hat{1}}_{-\hat{4}}+
 \Psi^{\hat{2}}_- \Psi^{\hat{4}}_- \omega^{\hat{1}}_{+\hat{4}}\right)+
  \left(\Psi^{\hat{1}}_+ \Psi^{\hat{3}}_+ \omega^{\hat{2}}_{-\hat{3}}+
 \Psi^{\hat{1}}_- \Psi^{\hat{3}}_- \omega^{\hat{2}}_{+\hat{3}}\right)\bigg)\label{2felorig}
\end{eqnarray}
and the order $\epsilon$ correction they yield is
\begin{eqnarray}
i\e \int d^2\sigma  &&\frac{1}{ \cosh^2\rho} \Big[ \tan\theta
\Big(\Psi^{\hat{3}}_+ \Psi^{\hat{4}}_+ \left(e^{2 i \omega - i
\varphi - i \phi_1} \bar \psi - e^{-2 i \omega + i \varphi + i
\phi_1} \bar \psi^\dagger\right)
\nonumber\\
&& +  \Psi^{\hat{3}}_- \Psi^{\hat{4}}_- \left(e^{2 i \omega - i
\varphi + i \phi_1} \psi - e^{-2 i \omega + i \varphi - i \phi_1}
\psi^\dagger\right) \Big)
\nonumber\\
&&  + 2 \tanh\rho \Big(\Psi^{\hat{2}}_+ \Psi^{\hat{4}}_+ e^{-3 i
\omega+i \phi_1} \bar \psi^\dagger+ \Psi^{\hat{1}}_+
\Psi^{\hat{3}}_+ e^{3 i \omega-i \phi_1} \bar\psi
\nonumber\\
&&  + \Psi^{\hat{2}}_- \Psi^{\hat{4}}_- e^{-3 i \omega-i \phi_1}
\psi^\dagger+ \Psi^{\hat{1}}_- \Psi^{\hat{3}}_- e^{3 i \omega+i
\phi_1} \psi\Big)\Big]\ . \label{2felorige}
 \end{eqnarray}

\no
In addition to those terms,
we obtain extra 2-fermi terms at order $\epsilon$ after applying (\ref{FR1}) to
 (\ref{boscorel}). They are
\begin{equation}
-i \e  \int d^2\sigma \frac{\tan\theta}{ \cosh^2\rho} \bigg(e^{2 i
\omega-i \varphi + i \phi_1}
 \psi  \Psi^{\hat{3}}_- \Psi^{\hat{4}}_- + e^{2 i \omega-i \varphi-i \phi_1 }
 \bar \psi \Psi^{\hat{3}}_+ \Psi^{\hat{4}}_+\bigg)+ {\rm c.c.}
 \end{equation}
 and therefore they cancel the terms in the first two lines of (\ref{2felorige}).
 All in all the order $\epsilon$ correction to the 2-fermion
 terms is
\ba \delta S_{2-{\rm fermi}}  = 2 i\e    \int  d^2\sigma &&
\frac{\sinh\rho}{ \cosh^3\rho}\Big(\Psi^{\hat{2}}_+
\Psi^{\hat{4}}_+ e^{-3 i \omega+i \phi_1} \bar \psi^\dagger +
\Psi^{\hat{2}}_- \Psi^{\hat{4}}_- e^{-3 i \omega-i \phi_1}
\psi^\dagger
\nonumber\\
&& + \Psi^{\hat{1}}_+ \Psi^{\hat{3}}_+ e^{3 i \omega-i \phi_1}
\bar \psi+\Psi^{\hat{1}}_- \Psi^{\hat{3}}_- e^{3 i \omega+i
\phi_1} \psi \Big)\ .\label{2fele}
 \ea
Again, taking into account \eqn{ginva} and \eqn{bosone} we see that this is the second line
of the perturbation
\eqn{petrre} based on CFT considerations, with the same infinitesimal parameter $\e$.

 \no
Finally we consider the 4-fermi terms. The original ones, using
the antisymmetry properties of the Riemann tensor to reduce the
number of terms, read:
\begin{eqnarray}
S_{4-{\rm fermi}}= 2  \int d^2\sigma &&\bigg[
R_{\hat{1}\hat{2}\hat{1}\hat{2}} \Psi^{\hat{1}}_+
\Psi^{\hat{2}}_+\Psi^{\hat{1}}_- \Psi^{\hat{2}}_- +
R_{\hat{3}\hat{4}\hat{3}\hat{4}} \Psi^{\hat{3}}_+
\Psi^{\hat{4}}_+\Psi^{\hat{3}}_- \Psi^{\hat{4}}_- +
 R_{\hat{2}\hat{4}\hat{2}\hat{4}}
\Psi^{\hat{2}}_+ \Psi^{\hat{4}}_+\Psi^{\hat{2}}_- \Psi^{\hat{4}}_-
\nonumber\\
&& + R_{\hat{1}\hat{3}\hat{1}\hat{3}} \Psi^{\hat{1}}_+
\Psi^{\hat{3}}_+\Psi^{\hat{1}}_- \Psi^{\hat{3}}_-
 + R_{\hat{2}\hat{4}\hat{3}\hat{4}}
\left(\Psi^{\hat{2}}_+ \Psi^{\hat{4}}_+\Psi^{\hat{3}}_-
\Psi^{\hat{4}}_- + \Psi^{\hat{3}}_+
\Psi^{\hat{4}}_+\Psi^{\hat{2}}_- \Psi^{\hat{4}}_-\right)
\nonumber\\
&&  +  R_{\hat{1}\hat{3}\hat{3}\hat{4}} \left(\Psi^{\hat{1}}_+
\Psi^{\hat{3}}_+\Psi^{\hat{3}}_- \Psi^{\hat{4}}_- +
\Psi^{\hat{3}}_+ \Psi^{\hat{4}}_+\Psi^{\hat{1}}_-
\Psi^{\hat{3}}_-\right)
\bigg]\ .
\end{eqnarray}
To order $\epsilon$ their contribution to the Lagrangian density is
 \begin{eqnarray}
\frac{2\e}{ k  \cosh^2\rho}  && \bigg[ -\tan^2\theta \cos
2(\omega-\varphi)
 \Psi^{\hat{3}}_+ \Psi^{\hat{4}}_+\Psi^{\hat{3}}_- \Psi^{\hat{4}}_- +
 \Big(
 \left(1-2 \sinh^2\rho\right) e^{-4 i \omega}
 \Psi^{\hat{2}}_+ \Psi^{\hat{4}}_+\Psi^{\hat{2}}_- \Psi^{\hat{4}}_- + {\rm c.c.}\Big)\nonumber\\
  && +  \Big(
 \tan\theta \tanh\rho e^{-3 i \omega + i \varphi}
 \left(\Psi^{\hat{2}}_+ \Psi^{\hat{4}}_+\Psi^{\hat{3}}_- \Psi^{\hat{4}}_- +
\Psi^{\hat{3}}_+ \Psi^{\hat{4}}_+\Psi^{\hat{2}}_-
\Psi^{\hat{4}}_-\right) + {\rm c.c.}\Big) \bigg]\ .
\label{4foriel}
\end{eqnarray}
We obtain also 4-fermion terms at the same order in $\epsilon$ by
applying (\ref{FR1}) to (\ref{boscorel}) and to (\ref{2felorig}).
That obtained from (\ref{boscorel})  read:\begin{equation} -2\e\
\int  d^2\sigma \frac{\tan^2\theta \cos 2(\omega-\varphi)}{k
\cosh^2\rho} \Psi^{\hat{3}}_+ \Psi^{\hat{4}}_+\Psi^{\hat{3}}_-
\Psi^{\hat{4}}_-\ ,
\end{equation}
whereas those coming from  (\ref{2felorig}) are
\begin{eqnarray}
2 \e\int d^2\sigma &&\bigg[ 2 \frac{\tan^2\theta \cos
2(\omega-\varphi)}{k \cosh^2\rho}
\Psi^{\hat{3}}_+ \Psi^{\hat{4}}_+\Psi^{\hat{3}}_- \Psi^{\hat{4}}_-\nonumber\\
&&-\frac{\tanh\rho \tan\theta}{ k \cosh^2\rho} e^{-3 i \omega + i
\varphi}\left( \Psi^{\hat{2}}_+ \Psi^{\hat{4}}_+\Psi^{\hat{3}}_-
\Psi^{\hat{4}}_- + \Psi^{\hat{3}}_+
\Psi^{\hat{4}}_+\Psi^{\hat{2}}_- \Psi^{\hat{4}}_-+{\rm
c.c.}\right)\bigg]\ .
\end{eqnarray}
Combining everything together we conclude that the total
deformation of the 4-fermion terms to leading order in $\epsilon$
is
\begin{equation}
\delta S_{4-{\rm fermi}}=2\e \int d^2\sigma  \frac{1-2 \sinh^2\rho}{k  \cosh^2\rho} \left(
e^{4 i \omega}
 \Psi^{\hat{1}}_+ \Psi^{\hat{3}}_+\Psi^{\hat{1}}_- \Psi^{\hat{3}}_-
+ e^{-4 i \omega} \Psi^{\hat{2}}_+ \Psi^{\hat{4}}_+\Psi^{\hat{2}}_- \Psi^{\hat{4}}_-\right)\ .
 \end{equation}
Using \eqn{ginva} and \eqn{bosone} we see that this is the last line
of the perturbation
\eqn{petrre} based on CFT considerations, with the infinitesimal parameter $\e$.

\no This completes our proof that the conjectured perturbation
\eqn{petrre} based on holography and exact CFT considerations
completely reproduces the one obtained from the $\s$-model
semiclassical approach.

\subsection{General deformations of circular NS5-brane distribution}

We analyze now general deformations of the circular NS5-brane system
corresponding to small changes of the radius $r_0 \rightarrow r_0+\delta r_0$.
We can expand $\delta r_0$ in Fourier modes $\delta r_0 = \epsilon r_0 \cos (n \psi)$
with $n \in \mathbb{Z}$ and focus on each mode separately. The corresponding
distribution describes NS5-branes smeared on the $x^8-x^9$ plane along the curve
\begin{equation}
\left(x^8\right)^2+\left(x^9\right)^2 = r^2_0+\xi(\psi)\ ,\qq
\psi=\tan^{-1}\left(\nicefrac{x^9}{x^8}\right) \ ,
\end{equation}
which represents a general planar deformation of the circular distribution. To linear
order in the deformation parameter we take
$\xi(\psi)=2 \epsilon \, r^2_0 \cos(n \psi)$ and
this is the continuous limit of the discrete distribution
(\ref{genvev1}) upon the identification $2(j+1)\equiv n $ and for
the values of $n$ where $0\leqslant j \leqslant (k-2)/2$. The mode with
$n=0$, i.e.~a uniform rescaling of the radius of the circle,
corresponds according to the discussion in Sec. \ref{sec:step222} to the
operator $j=(k-2)/2$. Since the latter is of order $k$ the
discussion here is not appropriate for that operator and
instead we refer the reader to Sec. \ref{sec:step222} where we
studied its semiclassical expression. The mode with $n=1$ is not
captured from the set of operators under consideration since the
corresponding value of $j=-\ha$ does not belong to the allowed
interval. In any case, we see that only a finite number of
continuous deformations have discrete analogues.
 These will be the deformations we will be interested in and hence we will
assume that $2\leqslant n\ll k$. The discrete distribution entails also
a non-trivial angular distribution $\lambda(\psi)$ and hence the
full NS5-brane density reads:
\begin{equation}
\rho({\bf x}) = \frac{1}{\pi} \lambda(\psi)
\delta\left(\left(x^8\right)^2+\left(x^9\right)^2-r^2_0-\xi(\psi)\right)
\delta\left(x^6\right)\delta\left(x^7\right)\ ,
\end{equation}
where $\l(\psi)$ is the angular density in \eqn{meas}. According
to this it is convenient to parameterize the Cartesian coordinates
along the curve as
\be x^8 = r_0 \cos\psi (1+\e \cos n \psi)\ ,\qq
x^9 = r_0 \sin\psi (1+\e \cos n \psi) \ ,
\label{hhho}
 \ee
 to linear order in
$\e$.\footnote{An alternative, equivalent to \eqn{jk99}, coordinate system is the one corresponding
to $ x^8 = r_0(\cos\phi +\e \cos (n-1) \phi) ,\ x^9 = r_0 (\sin\phi -\e \sin
(n-1) \phi)$,
with the angular relation, again to linear order in
$\e$, $\psi =\phi-\e\sin n \phi$. In this system the angular
distribution is easily seen to be uniform, i.e. $\l(\psi)d\psi =
d\phi$. This is the coordinate system used in \cite{Marios
Petropoulos:2005wu} for $n=2$.}
The deformation is geometrically
depicted in Figs. 1 and 2.
%%%%%%%%%%%%%%%%%%%%%%%%%%%%%%%%%%%%%%%%%%%%%%%%
\begin{figure}[!ht]%[h!]
\begin{center}
\includegraphics[height= 5 cm,angle=0]{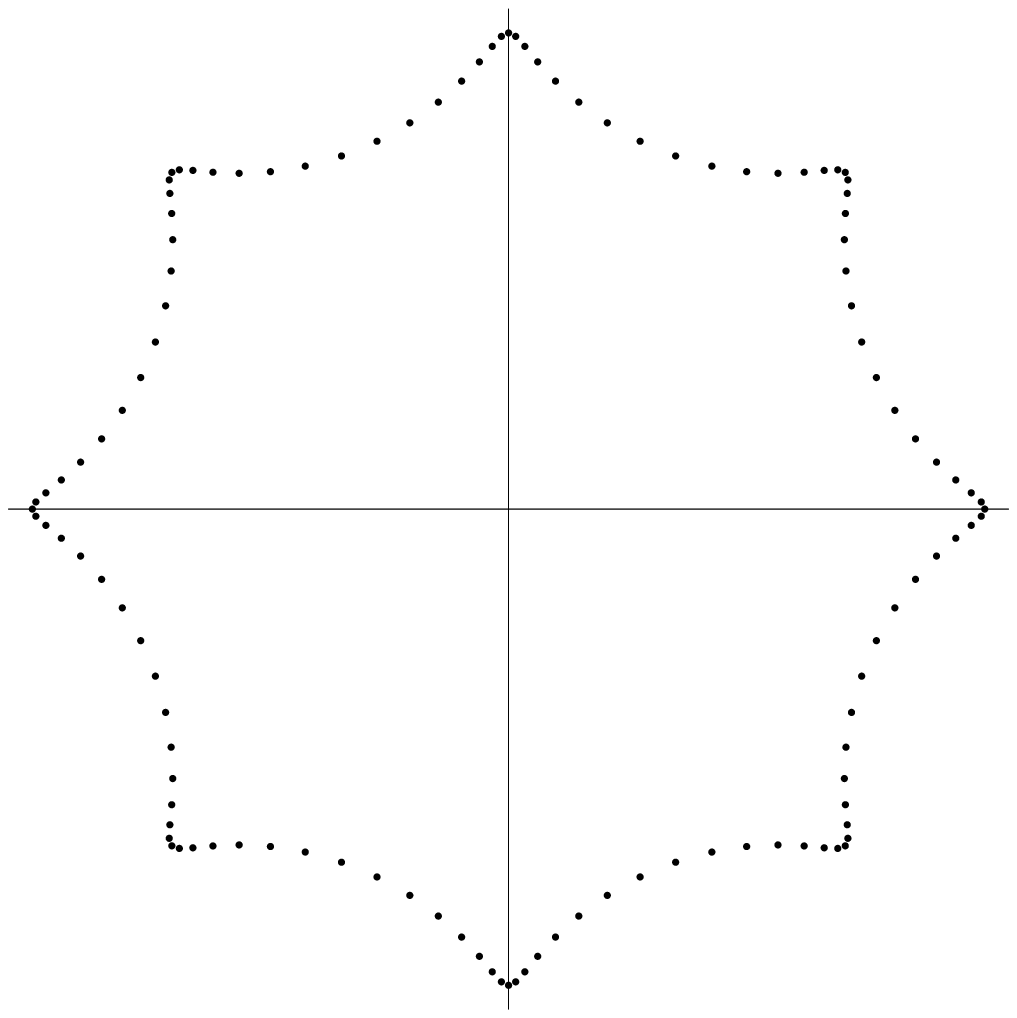}\hskip 2cm
\includegraphics[height= 5 cm,angle=0]{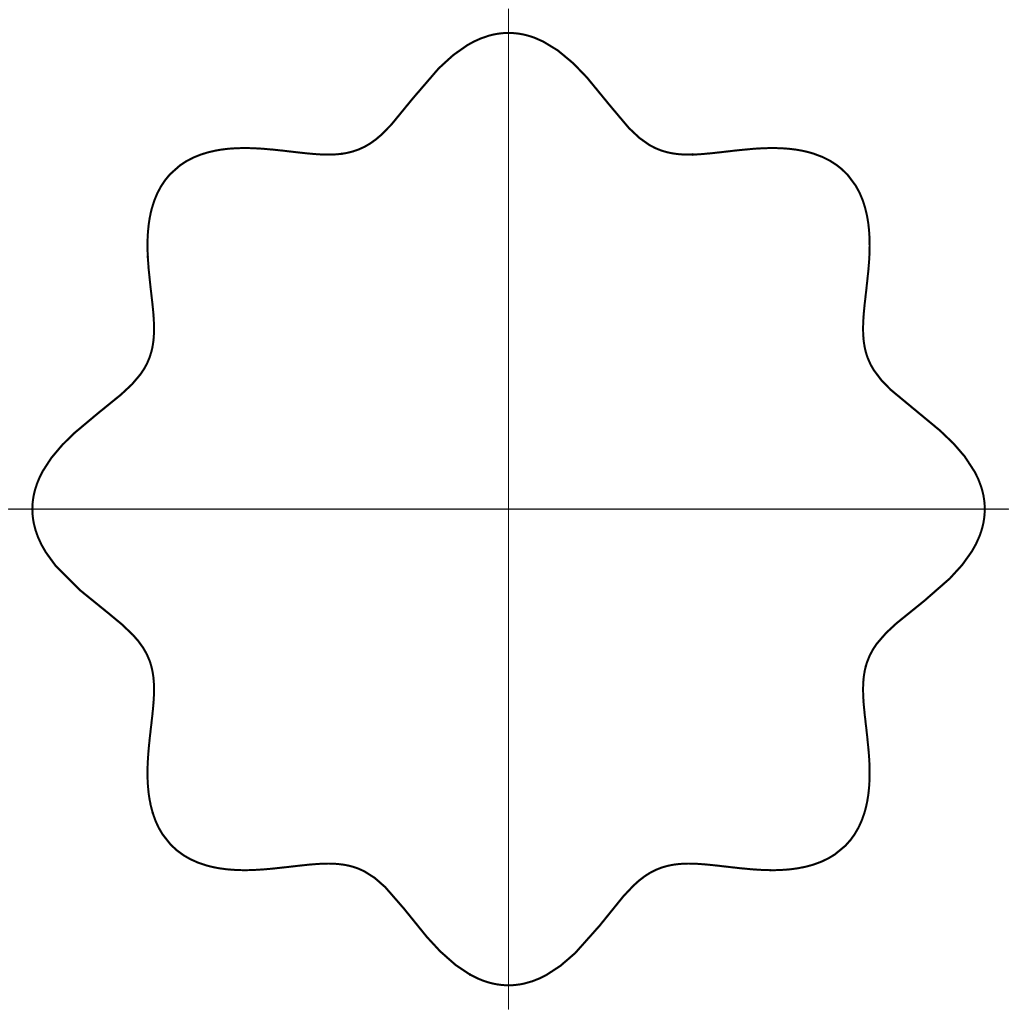}
\end{center}
\vskip -.5 cm \caption{On the left, the discrete distribution of
$k\sim 120$ branes on a deformed circle for $n=8$ $(j=3)$ and
$\e=0.1$. It is reproduced using \eqn{genvev1}, which in the
continuum is equivalent to the parametrization of the deformation described in footnote 6.
On the right, we depict the same deformation in the continuous approximation
using the parametrization \eqn{hhho}. }
\end{figure}
%%%%%%%%%%%%%%%%%%%%%%%%%%%%%%%%%%%%%%%%%%%%%%%%

 \no
Using the above, the harmonic function \eqn{haahm}, in the
near-horizon limit where the 1 is dropped, reads:
\begin{eqnarray}
H_n(\rho,\theta,\psi)=&&
\frac{k}{2\pi} \\
\int_0^{2 \pi}&& \!\!\!d\psi' \frac{\lambda(\psi')}
{r^2+r_0^2+r_0^2 \sin^2\theta+\xi(\psi')-2 r_0 \sqrt{r^2+r_0^2}
\sqrt{r_0^2+\xi(\psi')} \sin\theta \cos(\psi-\psi')}\nonumber .
\end{eqnarray}
We are interested in the leading order $\epsilon$ correction to $H$
which is given by
\begin{equation}
\delta H_n = \epsilon \frac{k}{2\pi} \int_0^{2\pi} d\psi'\Bigg(
 \frac{(n-1)\ \cos n \psi'}{A-B \cos(\psi'-\psi)}
+ \frac{(A-2 r_0^2) \cos n\psi'}{\big(A-B \cos(\psi'-\psi)\big)^2}
\Bigg)\ ,
\end{equation}
where we defined for convenience \be
A(\rho,\theta)=r^2+r_0^2+r_0^2\sin^2\theta\ ,\qq
 B(\rho,\theta)=2 r_0 \sqrt{r^2+r_0^2} \sin\theta\ .\nonumber
\ee
Shifting the integration variable as $\psi' \rightarrow \psi'+\psi$
enables us to write the previous expression
in terms of the integral
\begin{equation}
I_n(A,B)=\frac{1}{2\pi} \int_0^{2\pi} d\psi' \frac{\cos n \psi'}{A-B\cos\psi'}=
\frac{1}{\sqrt{A^2-B^2}}\Bigg(\frac{A-\sqrt{A^2-B^2}}{B}\Bigg)^n\ ,
\end{equation}
as
\begin{equation}
\delta H_n = \epsilon k \cos n \psi \Big((n-1) I_n(A,B)-(A-2 r_0^2) \frac{\partial}{\partial A}
I_n(A,B)\Big)\ ,
\end{equation}
where we have used the fact that the same integral as $I_n$  but with the
cosine replaced by sine vanishes due to its parity.
Explicitly, the final result is
\begin{equation}\label{gendefh}
\delta H_n  = \epsilon k \cos n\psi \Bigg(\frac{r_0
\sin\theta}{\sqrt{r^2+r_0^2}}\Bigg)^n \, \frac{n r^2
(2r^2+r_0^2)-r_0^4
-r_0^2\big((2-n)r^2+r_0^2\big)\cos2\theta}{(r^2+r_0^2\cos^2\theta)^3}\
.
\end{equation}
The deformation of the harmonic function above yields immediately the corresponding
deformation of the geometry of the transverse space and the full metric becomes
\begin{eqnarray}
\frac{ds^2}{k} &=& \left( \frac{dr^2}{r^2+r^2_0} +
d\theta^2 + \frac{r^2\cos^2 \theta
d\tau^2}{r^2+r^2_0\cos^2 \theta}
+\frac{\left(r^2+r^2_0\right)\sin^2 \theta
d\psi^2}{r^2+r^2_0\cos^2 \theta}
\right)\bigg[1+\epsilon  \times \nonumber \\
&&\times \label{totdef} \cos n\psi \Bigg(\frac{r_0
\sin\theta}{\sqrt{r^2+r_0^2}}\Bigg)^n \, \frac{n r^2 (2r^2+r_0^2)
-r_0^4
-r_0^2\big((2-n)r^2+r_0^2\big)\cos2\theta}{(r^2+r_0^2\cos^2\theta)^3}\bigg]
\ .
\end{eqnarray}
Recall, however, that the holographic approach refers to the
T-dual of the original geometry and it is necessary to know also
the deformation of the $B$ field in order to T-dualize. Instead of
following this route, we will start from the holographic result
for the bosonic deformation and show that after T-duality and
appropriate reparameterization yields exactly the deformation of
the transverse metric induced by (\ref{gendefh}). Subsequently, we
will verify that the fermionic deformations of the ${\cal N}=1$
$\sigma$ model match the holographic predictions.

\no
The first line of \eqn{hologenop} is the purely bosonic deformation of the $SL(2,\mathbb{R})/U(1) \times
SU(2)/U(1)$ theory
\begin{equation}
\delta S_{\rm bose} = \epsilon \int d^2\sigma   \frac{k \sin^{n-2}\theta}{\cosh^n\rho}
\left(e^{i (n \omega-(n-2)\varphi)}\psi \bar \psi + {\rm c.c.} \right)\ ,
\label{boscorgen}
\end{equation}
 which implies a corresponding deformation of the metric
\begin{equation}
ds_n^{2} = 2 \epsilon k  \frac{\sin^{n-2}\theta}{\cosh^{n}\r}
\big[ \cos n(\omega -\varphi)  \left(d\theta^2-\tan^2\theta
d\varphi^2\right)+ 2 \sin n(\omega-\varphi) \tan\theta d\varphi
d\theta\big]\ .
\end{equation}
Explicitly the full metric becomes
\begin{eqnarray}
 {ds^{2}\ov k}  &=&
 d\rho^2+\coth^2\rho \, d\om^2
+d\th^2 + \tan^2\th \, d\varphi^2
+2 \epsilon \frac{\sin^{n-2}\theta}{\cosh^{n}\r}
\times
\nonumber \\
&& \times\big( {\cos n(\omega-\varphi)} \left(d\th^2-\tan^2\th \,
d\varphi^2\right)+2\sin n(\omega-\varphi) \tan\th\, d\varphi \,
d\th \big)\ .\label{js79gen}
\end{eqnarray}
The dilaton is not modified and hence we can read it from
(\ref{sigmacirclecft})
\begin{equation}\label{dilgen}
 e^{-2\Phi} = \sinh^2 \rho \cos^2 \theta\ ,
\end{equation}
while the antisymmetric tensor remains vanishing. As a first check one can indeed
verify that the above background is a solution of the equations of motion
to leading order in $\epsilon$.

\no
We proceed now with the T-duality. For that purpose we need to
 perform as before the reparameterizations
$\omega=\psi+\tau$ and $\varphi=\tau$ which make manifest the isometry
corresponding to shifts of the new coordinate $\tau$. T-dualizing along $\partial_\tau$
yields
a new metric
\begin{eqnarray}\label{tdualjdef}
\frac{ds^2}{k}&=&  \frac{dr^2}{r^2+r^2_0} +
d\theta^2 + \frac{r^2\cos^2 \theta
d\tau^2}{r^2+r^2_0\cos^2 \theta}
+\frac{\left(r^2+r^2_0\right)\sin^2 \theta
d\psi^2}{r^2+r^2_0\cos^2 \theta}
\nonumber\\
&&+2 \epsilon
\Bigg(\frac{r_0 \sin\theta}{\sqrt{r^2+r_0^2}}\Bigg)^n
\Bigg\{ -\sin n \psi \sin 2 \theta
\frac{\left(r^2+r^2_0\right)d\theta
d\psi}{r^2+r^2_0\cos^2 \theta}
\nonumber\\
&&+ \cos  n \psi \left(d\theta^2 +\left(\frac{ \sin\theta
\cos\theta}{r^2+r^2_0\cos^2 \theta}\right)^2 \left( r^4
d\tau^2 - \left(r^2+r^2_0\right)^2 d\psi^2\right)
\right)\Bigg\}\ ,
\ea
a dilaton
\be
 e^{-2\Phi} = \frac{r^2+r^2_0\cos^2\theta}{r^2_0} -
2 \epsilon \cos  n \psi \frac{r^2}{r^2_0}
\Bigg(\frac{r_0 \sin\theta}{\sqrt{r^2+r_0^2}}\Bigg)^n
\ee
and a two-form potential with non-vanishing components
\begin{eqnarray}
\frac{B_{\tau\psi}}{k}&=&\frac{1}{\Sigma} +
\epsilon \cos  n \psi\frac{ r^2 }{\Sigma^2
\cos^2 \theta }\frac{r^{n}_0
\sin^{n}\theta}{\left(r^2+r^2_0\right)^{\frac{n}{2}+1}}\ ,
\nonumber\\
\frac{B_{\tau\theta}}{k}&=& \epsilon \sin  n
\psi \frac{ r^2 }{\Sigma \sin\theta \cos \theta }\frac{r^{n}_0
\sin^{n}\theta}{\left(r^2+r^2_0\right)^{\frac{n}{2}+1}}\ ,
\end{eqnarray}
where we have defined
\begin{equation}
\Sigma = 1 + \frac{r^2 \tan^2\theta}{r^2+r_0^2}\ .
\end{equation}
 Finally, if we perform the following reparameterizations
\begin{eqnarray}
\delta r &=& - {\epsilon r\ov n-1}
\frac{r^{n}_0\sin^{n}\theta}{\left(r^2+r^2_0\right)^{\frac{n}{2}-1}\left(r^2+r^2_0\cos^2\theta\right)}
\cos n\psi\ ,
\nonumber\\
\delta \theta &=& -{ \epsilon \cot\theta\ov n-1}
\frac{r^{n}_0\sin^{n}\theta}{\left(r^2+r^2_0\right)^{\frac{n}{2}-1}\left(r^2+r^2_0\cos^2\theta\right)}
\cos n\psi\ ,
\\
\delta  \psi &=&  {\epsilon\ov n-1}
\frac{1}{\sin^2\theta}
\frac{r^{n}_0\sin^{n}\theta}{\left(r^2+r^2_0\right)^{\frac{n}{2}}} \sin
n\psi\ ,
\nonumber
\end{eqnarray}
and set $r=r_0\sinh\r$, we find that the metric in the new
coordinate system is exactly the metric (\ref{totdef}) we found
earlier. Note that these reparameterizations do not change the
angular character of the variables $\th$ and $\psi$ since they are
nowhere singular and their periodicity is respected.

\no We proceed now to the analysis of the fermionic terms in the
${\cal N}=1$ $\sigma$ model. The necessary spin-connection and
curvature two-form associated to the metric (\ref{js79gen}) are
given in App. \ref{geodata}. As in the case of the elliptic
deformation, we need first perform the redefinitions (\ref{FR1})
in order to render the $\sigma$-model fermions free. The original
order $\epsilon$ 2-fermi terms are
\begin{eqnarray}
i \epsilon \int  d^2\sigma  &&\frac{ \sin^{n-2}\theta}
{\cosh^n\rho}
 \bigg[ \tan\theta \Big(\Psi^{\hat{3}}_+ \Psi^{\hat{4}}_+
\left(e^{n i( \omega-\varphi) + i \varphi - i \phi_1} \bar \psi -
e^{-n i (\omega-\varphi) - i \varphi + i \phi_1} \bar
\psi^\dagger\right)
\nonumber\\
&& +  \Psi^{\hat{3}}_- \Psi^{\hat{4}}_- \left(e^{n i
(\omega-\varphi) + i \varphi + i \phi_1} \psi - e^{-n i( \omega
-\varphi)- i \varphi - i \phi_1}  \psi^\dagger\right)\Big)
\nonumber\\
&& + n \tanh\rho \Big(\Psi^{\hat{2}}_+ \Psi^{\hat{4}}_+
 e^{-n i (\omega-\varphi)-2i \varphi - i \omega+i \phi_1} \bar
\psi^\dagger+ \Psi^{\hat{1}}_+ \Psi^{\hat{3}}_+ e^{n i
(\omega-\varphi)+2i \varphi + i \omega-i \phi_1} \bar\psi
\nonumber \\
&& + \Psi^{\hat{2}}_- \Psi^{\hat{4}}_- e^{-n i (\omega-\varphi)-2i
\varphi - i \omega-i \phi_1} \psi^\dagger+ \Psi^{\hat{1}}_-
\Psi^{\hat{3}}_- e^{n i (\omega-\varphi)+2i \varphi + i \omega+i
\phi_1} \psi\Big)\bigg]\ \label{2fgenorige}
 \end{eqnarray}
and, along with the contribution of (\ref{boscorgen}) from the field
redefinition \eqn{FR1}, combine to
 \ba
\delta S_{2-{\rm fermi}}   = i \epsilon n
  \int  d^2\sigma  &&  \frac{ \sin^{n-2}\theta \sinh\rho}{ \cosh^3\rho}
\Big(\Psi^{\hat{2}}_+ \Psi^{\hat{4}}_+
  e^{-n i (\omega-\varphi)-2i \varphi - i \omega+i \phi_1}
\bar \psi^\dagger
\nonumber\\
&&+ \Psi^{\hat{2}}_- \Psi^{\hat{4}}_- e^{-n i (\omega-\varphi)-2i
\varphi - i \omega-i \phi_1} \psi^\dagger
  + \Psi^{\hat{1}}_+ \Psi^{\hat{3}}_+ e^{n i (\omega-\varphi)+2i \varphi
 + i \omega-i \phi_1} \bar\psi
\nonumber\\
&& + \Psi^{\hat{1}}_- \Psi^{\hat{3}}_- e^{n i (\omega-\varphi)+2i
\varphi + i \omega+i \phi_1} \psi \Big)\ . \label{2fgen}
 \ea
This matches the operators in the second line of (\ref{hologenop})
that we computed using the conjectured holography and CFT.

\no Let us finally check the 4-fermion terms. Besides the ones
coming from the original 4-fermi interactions we have additional
ones coming from (\ref{boscorgen})  and from the original
2-fermion terms after performing the field redefinition
(\ref{FR1}). Combining everything together results in the
expression \ba \delta S_{4-{\rm fermi}} =   \e n  \int  d^2\sigma
&& \frac{(1-n \sinh^2\rho)\sin^{n-2}\theta}{k  \cosh^n\rho} \Big(
e^{i n (\omega-\varphi)+ 2 i \omega+2 i \varphi}
 \Psi^{\hat{1}}_+ \Psi^{\hat{3}}_+\Psi^{\hat{1}}_- \Psi^{\hat{3}}_-
\nonumber\\
&&+ e^{-i n (\omega-\varphi)- 2 i \omega-2 i \varphi}
\Psi^{\hat{2}}_+ \Psi^{\hat{4}}_+\Psi^{\hat{2}}_-
\Psi^{\hat{4}}_-\Big)\ ,
 \ea
which matches exactly the last line of (\ref{hologenop}).

\section{Summary and perspectives}

The starting point of our analysis was an exact $\mathcal{N}=4$
string background (metric, antisymmetric tensor and dilaton),
generated by $k$ parallel NS5-branes distributed over a circle
transverse to their worldvolume. In the near-horizon regime it is described
by the Kazama--Suzuki model
$SU(2)/U(1) \times SL(2,\mathbb{R})/U(1)$.

\no The geometric deformations of this setup can be studied in two
different manners. Either by using holography, which relates
NS5-brane worldvolume operators to $\sigma$-model worldsheet
operators (Sec. \ref{sec:step2}), or by explicitly perturbing the
locus of the NS5-branes, from a circle to an $n$-bump deformed
circle, and computing the corresponding $\sigma$-model background
fields  (Sec. \ref{sec:step3}). From the latter we can read-off
the worldsheet operators that trigger the perturbation. These are
marginal since the NS5-brane perturbations do not alter the
conformality of the $\mathcal{N}=4$ string backgrounds at hand.
Actually, they exactly marginal as they originate from chiral primary operators.
They are indeed described in terms of compact parafermions dressed
with non-compact and compact primaries, whose conformal weights
add up to $(1,1)$.
All these generalize the results obtained in \cite{Marios
Petropoulos:2005wu} for the elliptical deformation of the circle,
to the infinite tower of $n$-bump modes. As already stressed,
these computations are performed in the semiclassical
approximation, but the whole framework allows to safely conjecture
their validity beyond that level.

\no The marginal worldsheet operators that are revealed by the
deformed-$\sigma$-model approach turn out to be in agreement with
those found using the LST dictionary for $n$-bump NS5-brane
displacements away from the circle. This brilliantly demonstrates
the validity of the holographic duality for a large class of
operators. To the best of our knowledge, this result is unique as
is the construction of the set of non-left-right-factorized
marginal operators of the $SU(2)/U(1) \times
SL(2,\mathbb{R})/U(1)$ conformal model.

\no One could generalize this study in a  variety of directions.
For instance, one can consider non-planar deformations of the
NS5-branes and verify the validity of the holographic dictionary.
Furthermore, it would be very interesting to extend this analysis
to time-dependent deformations where issues  like supersymmetry
breaking can be addressed in an exact CFT framework. A challenging
open problem pertains as to the exact CFT description of NS5-brane
systems, for instance the elliptic distribution, which preserve a
large part of the original transverse-space symmetries. Presumably
generalizations of the compact and non-compact parafermions would
be instrumental for such an endeavor.

\section*{Acknowledgements}

The authors would like to thank their colleagues C.~Bachas,
C.~Kounnas and  V.~Niarchos for stimulating exchanges. The work of
Angelos Fotopoulos is partially supported by the European
Community's Human Potential Programme under contract
MRTN-CT-2004-005104 and by the Italian MUR under contracts
PRIN-2005023102 and PRIN-2005024045. Angelos Fotopoulos would like
to thank CERN Theory Division for its warm hospitality where part
of this work has been done. Marios Petropoulos thanks Neuch\^atel
University, Patras University and CERN Theory Division for kind
hospitality at various stages of this collaboration, and
acknowledges financial support by the Swiss National Science
Foundation, the French Agence Nationale pour la Recherche, contract 05-BLAN-0079-01,
and the
EU under the contracts MEXT-CT-2003-509661,  MRTN-CT-2004-005104
and MRTN-CT-2004-503369. Nikolaos Prezas wishes to thank the Ecole
Polytechnique for its warm hospitality where part of this work was
done. Konstadinos Sfetsos acknowledges partial support provided
through the European Community's program ``Constituents,
Fundamental Forces and Symmetries of the Universe'' with contract
MRTN-CT-2004-005104, the INTAS contract 03-51-6346 ``Strings,
branes and higher-spin gauge fields'' and the Greek Ministry of
Education programs $\Pi \Upsilon \Theta \mathrm{A} \Gamma
\mathrm{OPA}\Sigma$ with contract 89194.

\appendix

\section{\boldmath  Linear-dilaton background and the $\mathcal{N}=4$
superconformal  algebra   \unboldmath}\label{CHS}

\no
The CHS conformal field theory  $\mathbb{R}_{\phi}\times SU(2)_k$
contains a linear dilaton along the $\phi$ direction with
background charge $q=\sqrt{\frac{2}{k}}$ (where $k$ is the number
of NS5-branes) and a ${\cal N}=1$
supersymmetric $SU(2)$ WZW model at level $k$.
We will exhibit the ${\cal N}=4$ superconformal algebra \cite{Ademollo:1976wv}
 supported on
 $\mathbb{R}_{\phi}\times SU(2)_k$ following the discussion of \cite{Aharony:2003vk}.

\no The supersymmetric WZW model is decomposed into a bosonic
$SU(2)$ WZW model at level $k-2$ with affine currents $J^i$ and
three free fermions $\psi_i, \; i=1,2,3.$ We denote by $\phi$ the
boson corresponding to the dilaton direction  and by $\psi_\phi$
is its superpartner. The operator algebra is
\begin{equation}
\label{N4alg}
\begin{array}{rcl}
\displaystyle{\partial\phi(z)\partial\phi(w)}&=&
\displaystyle{-\frac{1}{(z-w)^2}} \ , \crbig
\displaystyle{\psi_i(z)\psi_j(w)} &=&
\displaystyle{\frac{\delta_{ij}}{z-w}} \ ,
    \crbig
\displaystyle{J^i(z)J^j(w)} &=&
\displaystyle{\frac{(k-2)\delta_{ij}}{2 (z-w)^2}+i \epsilon^{ijk}
\frac{J^k(w)}{z-w}} \ ,
\end{array}
\end{equation}
with $\epsilon^{123}=1$. We can change basis to $J^{\pm}=J^1\pm i
J^2$ and the corresponding OPEs read:
\begin{equation}
\label{OPEJ}
\begin{array}{rcl}
\displaystyle{J^3(z)J^3(w)}&=&\displaystyle{\frac{k-2}{2 (z-w)^2}}
\ , \crbig \displaystyle{J^3(z)J^\pm(w)}&=&\displaystyle{\pm
\frac{J^\pm(w)}{z-w}} \ , \crbig
\displaystyle{J^+(z)J^-(w)}&=&\displaystyle{\frac{k-2}{(z-w)^2}+\frac{2J^3(w)}{z-w}}
\ .
\end{array}
\end{equation}
We now define
\begin{eqnarray}\label{psidef}
\psi^\pm&=&\frac{1}{\sqrt{2}}(\psi_1\pm i \psi_2)\ ,\\
\psi&=&\frac{1}{\sqrt{2}}(\psi_\phi+i \psi_3)
\end{eqnarray}
and their OPEs read:\begin{equation} \label{OPEpsi}
\psi(z)\psi^*(w)=\psi^+(z)\psi^-(w)=\frac{1}{z-w}\ .
\end{equation}
The currents of the supersymmetric $SU(2)$ WZW model at level $k$
are given by
\begin{equation}\label{SUcur}
J_i^{\rm tot}=J_i-\frac{i}{2}\epsilon_{ijk} \psi_j \psi_k
\end{equation}
and in particular
\begin{equation}\label{J3SU}
J_3^{\rm tot}=J_3-i \psi_1\psi_2=J_3+\psi^+\psi^-\ .
\end{equation}
The primaries $\Phi_{j;m,\bar m}^{ su}$ of the bosonic $SU(2)$ WZW
model have conformal weight
\begin{equation}\label{SUpricw}
h=\frac{j(j+1)}{k}\ ,
\end{equation}
while those of the linear dilaton theory have
\begin{equation}\label{LDpricw}
h\left(e^{a \phi}\right)=-\frac{1}{2} a (a+q)\ .
\end{equation}

\no
The system $\mathbb{R}_{\phi}\times SU(2)$ is known to exhibit
a ``small" ${\cal N}=4$ superconformal symmetry, in line with the
fact that the dual configuration of NS5-branes is 1/2 BPS and in
type II string theories preserves 16 supersymmetries in
space--time. The superconformal generators are
\begin{equation}
\label{Ggen}
\begin{array}{rcl}
 \displaystyle{G}&=&\displaystyle{i \psi_\phi \partial\phi +i q
 \partial\psi_\phi+ q(J_1 \psi_1 + J_2 \psi_2+ J_3 \psi_3 -i \psi_1
 \psi_2 \psi_3)} \ , \crbig
 \displaystyle{G_1}&=&\displaystyle{i \psi_1 \partial\phi +i q \partial\psi_1+ q(-J_1
\psi_\phi + J_2 \psi_3- J_3 \psi_2 +i \psi_2 \psi_3\psi_\phi)} \ ,
 \crbig
 \displaystyle{G_2}&=&\displaystyle{i \psi_2 \partial\phi +i q \partial\psi_2+
q(-J_2 \psi_\phi + J_3 \psi_1- J_1 \psi_3 +i \psi_3 \psi_1
\psi_\phi)} \ , \crbig
 \displaystyle{G_3}&=&\displaystyle{i \psi_3 \partial\phi +i q\partial\psi_3+ q(-J_3 \psi_\phi +
J_1 \psi_2- J_2 \psi_1 +i \psi_1 \psi_2 \psi_\phi)}
\end{array}
\end{equation}
and the $SU(2)$ $R$-symmetry currents are
\begin{equation}
\label{Sgen}
\begin{array}{rcl}
 \displaystyle{S_1}&=&\displaystyle{-\frac{i}{2}(\psi_\phi \psi_1+\psi_2\psi_3)} \ , \crbig
 \displaystyle{S_2}&=&\displaystyle{-\frac{i}{2}(\psi_\phi \psi_2+\psi_3\psi_1)} \ , \crbig
 \displaystyle{S_3}&=&\displaystyle{-\frac{i}{2}(\psi_\phi \psi_3+\psi_1\psi_2)} \ .
 \end{array}
\end{equation}
These currents generate an $SU(2)$ current algebra at level one.
The energy--momentum tensor reads:
\begin{equation}\label{Tgen}
T=-\frac{1}{2}(\partial\phi)^2-\frac{1}{2} q
\partial^2\phi+\frac{J^i J^i}{k}-
\frac{1}{2}\psi^*\partial\psi-\frac{1}{2}\psi\partial\psi^*-\frac{1}{2}\psi^+\partial\psi^-
-\frac{1}{2}\psi^-\partial\psi^+\ .
\end{equation}

\no
 For our purposes it will be useful to exhibit an ${\cal N}=2$ subalgebra
 of the above algebra with $G$ and $G_3$ the corresponding superconformal
 generators.  We can define
 \begin{equation}\label{N4Gpm}
 G^\pm=\frac{1}{\sqrt{2}}\left(G\pm i G_3\right)
 \end{equation}
 and explicitly they are given by\footnote{An interesting feature of this construction is that
 although $\partial\phi$ is not a conformal primary field in the linear-dilaton
 background, the $G^\pm$ are primaries since the cubic term
 coming from contracting $\partial^2\phi $ with $\partial\phi$ cancels out another cubic term coming from $\psi\partial\psi^*$ or $\psi^*\partial\psi$
 contracted with either $\partial\psi$ or $\partial\psi^*$.}
 \begin{eqnarray}\label{CHSGpm2}
 G^+&=&i\psi\left(\partial\phi- q J_3^{\rm tot}\right)+i q \partial\psi+ q J^-\psi^+\ ,\\
 G^-&=&i\psi^*\left(\partial\phi+ q J_3^{\rm tot}\right)+i q \partial\psi^*+ q J^+\psi^-\ .
 \end{eqnarray}
 The $U(1)$ $R$-symmetry generator is
 \begin{equation}\label{CHSJR}
 J_{\mathrm{R}}=\psi\psi^*+\psi^+\psi^-=2 S_3
 \end{equation}
and the level one $SU(2)$ raising and lowering operators
 are
 \begin{equation}\label{Spm}
S^+= S_2+ i S_1=  \psi \psi^+\ , \qquad S^-=S_2-iS_1=  \psi^-
\psi^*\ .
\end{equation}
We can also form the combinations
 \begin{equation}\label{N4tGpm}
 \tilde G^\pm=\frac{1}{\sqrt{2}}\left(G_1\pm i G_2\right)\ ,
 \end{equation}
 which explicitly read:
\begin{equation}
\label{CHStGpm2}
\begin{array}{rcl}
\displaystyle{\tilde G^+}&=&\displaystyle{i
\psi^+\big(\partial\phi+q (J_3-\psi\psi^*)\big)+i q
\partial\psi^+-
 q J^+\psi} \ , \crbig
\displaystyle{\tilde G^-}&=&\displaystyle{i
\psi^-\big(\partial\phi-q (J_3-\psi\psi^*)\big)+i q
\partial\psi^--
 q J^-\psi^* } \ .
\end{array}
\end{equation}

\no
The operators defined above satisfy the small ${\cal N}=4$  superconformal algebra
\begin{equation}
\label{OPE1}
\begin{array}{rclrcl}
\displaystyle{J_{\mathrm{R}}(z) S^\pm (w)}&\sim &\displaystyle{\pm
\frac{2 S^\pm}{z-w}} \ , &\quad
 \displaystyle{S^+(z) S^-(w) }&\sim
 &\displaystyle{\frac{J_{\mathrm{R}}(w)}{z-w} + \frac{1}{(z-w)^2}}
 \ , \crbig
 \displaystyle{S^+(z) G^-(w)}&\sim &\displaystyle{- \frac{\tilde{G}^+}{z-w}}\ , &\quad
 \displaystyle{S^+(z)
\tilde{G}^-(w)}&\sim &\displaystyle{ \frac{G^+}{z-w}} \ , \crbig
 \displaystyle{S^-(z) G^+(w)}&\sim&\displaystyle{- \frac{\tilde{G}^-}{z-w}}\ , &\quad
 \displaystyle{S^-(z)\tilde{G}^+(w) }&\sim&\displaystyle{ \frac{G^-}{z-w}} \
 .
\end{array}
\end{equation}
The OPEs of $G^\pm$ and $\tilde{G}^\pm$ with $J_{\mathrm{R}}$
imply the charges $+1$ for $G^+, \tilde{G}^+$ and $-1$ for $G^-,
\tilde{G}^-$. The remaining OPEs of the $SU(2)$ currents with the
supercharges are regular. In addition the doublets have singular
OPEs among themselves
\begin{equation}
\label{OPE3}
\begin{array}{rcl}
 \displaystyle{G^+(z) G^-(w) }&\sim &\displaystyle{\frac{2 c
}{3(z-w)^3} + \frac{2 J_{\mathrm{R}}(w)}{(z-w)^2} + \frac{2T(w) +
\partial J_{\mathrm{R}}(w)}{z-w}} \ , \crbig
 \displaystyle{\tilde{G}^+(z) \tilde{G}^-(w) }&\sim
 &\displaystyle{\frac{2 c }{3(z-w)^3} + \frac{2
J_{\mathrm{R}}(w)}{(z-w)^2} + \frac{2T(w) + \partial
J_{\mathrm{R}}(w)}{z-w}} \ , \crbig
 \displaystyle{G^+(z) \tilde{G}^+(0)}&\sim &\displaystyle{\frac{S^+(w)}{(z-w)^2} +
\frac{\partial S^+(w)}{2(z-w)}}
 \ , \crbig
 \displaystyle{G^-(z) \tilde{G}^-(0) }&\sim
 &\displaystyle{\frac{S^-(w)}{(z-w)^2} + \frac{\partial S^-(w)}{2(z-w)}} \ ,
\end{array}
\end{equation}
where $c=6$ for the CHS background.

\no
 In the remaining part of this appendix we would like to
 demonstrate that the operators in (\ref{LGW2}), when inserted in (\ref{N=4pert}), leave the
 ${\cal N}=4$
 algebra unbroken. We follow the analysis of \cite{Berkovits:1994vy}. As it is
 well-known a perturbation which preserves at least an ${\cal N}=2$ should be constructed using chiral or
 antichiral fields. The chiral fields $O_a$ are annihilated by
 $G^+_{-\frac{1}{2}}$ and have charge ${\cal Q}_{\mathrm{R}}= +1$ and dimension $h={1\over
 2}$. The antichiral fields are annihilated by $G^-_{-{1\over 2}}$ and have charge
 ${\cal Q}_{\mathrm{R}}=-1$
 and dimension $h={1\over 2}$.
 The general perturbation which preserves the ${\cal N}=2$ subalgebra
 containing  $G^-$ and $ G^+$ is
 \begin{equation}\label{N=2pert}
 \delta S= \int d^2z \; \big(\lambda_a G^- O_a + \bar\lambda_a G^+ \bar
 O_a\big)
\end{equation}
where $O_a/\bar O_a$ is an chiral/antichiral field and $G^- O_a$
means $\oint dz G^-(z) O_a (0)$ (which will be implied for the
rest of this appendix). The index $a$ parameterizes the set of
chiral (antichiral) operators. Then it turns out that the
requirement of ${\cal N}=4$ SCFT invariance imposes the additional
constraints:
\begin{equation}\label{N=4con}
S^- O_a = M_a^{\bar b } \bar O_a\ , \qquad S^+ \bar O_a = -M_{\bar
a}^{ b} O_a\ , \ \ M_a^{\bar b} M_{\bar b}^{ c}= \delta_a^c \ .
\end{equation}

\no Now we will consider the operators (\ref{LGW2}). In this case
the matrix $M_a^{\bar b}$ is  one by one and can be set to unity.
Therefore  the index $a$ can be dropped and the operator under
consideration is
\begin{equation}\label{O}
O= \psi^+ \Phi^{{ su}}_{j;j,j} e^{-q (j+1) \phi} \ ,
\end{equation}
where we write only its holomorphic part. It is easy to verify, using the explicit
expressions of the
${\cal N}=4$ generators, that it is indeed a chiral operator as claimed in the
main text. Moreover, it is straightforward to show that the operator
\begin{equation}\label{bO}
\bar O= S^- O= -\psi^* \Phi^{{ su}}_{j;j,j} e^{-q (j+1) \phi}
\end{equation}
is an antichiral operator. Hence the constraints (\ref{N=4con}) are satisfied
and perturbing the theory with the upper component of any of the operators
(\ref{O}) does not spoil ${\cal N}=4$ superconformal invariance.
The same discussion can be trivially extended to the antiholomorphic sector.

\section{\boldmath Parafermionic operator products   \unboldmath}\label{pOPE}

\subsection{Compact parafermions and $SU(2)$ current algebra}

Let us consider the  $SU(2)$ current algebra at level $k\geq 2$ whose currents
obey the following operator algebra:
\begin{equation}
\label{SUalg}
\begin{array}{rcl}
\displaystyle{J^3(z)
J^3(w)}&\sim&\displaystyle{\frac{k}{2}\frac{1}{(z-w)^2}} \ ,
\crbig \displaystyle{J^3(z)J^\pm(w)}&\sim&\displaystyle{\pm
\frac{J^\pm(w)}{z-w}} \ , \crbig
\displaystyle{J^+(z)J^-(w)}&\sim&\displaystyle{\frac{k}{(z-w)^2}+
\frac{2 J^3(w)}{z-w}}
\end{array}
\end{equation}
and decompose the currents as
\begin{equation}
\label{su2c2}
\begin{array}{rcl}
\displaystyle{J^3}&=&\displaystyle{i\sqrt{\frac{k}{2}} \partial P} \
, \crbig \displaystyle{J^{+}}&=&\displaystyle{\sqrt{k} \psi \exp
i\sqrt{\frac{2}{k}}P} \ , \crbig
\displaystyle{J^{-}}&=&\displaystyle{\sqrt{k} \psi^\dagger \exp
-i\sqrt{\frac{2}{k}}P } \ ,
\end{array}
\end{equation}
with $P$ being a free boson and $\psi,\psi^\dagger$ being the basic the parafermion fields.

\no In general the compact parafermionic algebra at level $k$
contains a set of objects $\psi_l$ and $\psi_l^\dagger=\psi_{-l}$
where $l=0,1,\ldots,k-1$ and so that $\psi_0=1$ and $\psi_1=\psi$
(see for instance \cite{Bakas:1991fs}). Their conformal dimensions
are $\Delta_{l}=l-\frac{l^2}{k}$. The generic parafermion OPEs are
\begin{equation}
\psi_{l_1}(z)\psi_{l_2}(w)=C_{l_1,l_2}
(z-w)^{\Delta_{l_1+l_2}-\Delta_{l_1}-\Delta_{l_2}}
\big(\psi_{l_1+l_2}(z) + {\cal O}(z-w)\big)\label{gcpope}
\end{equation}
and
\begin{equation}
\psi_{l}(z)\psi_{l}^\dagger(w)=(z-w)^{-2\Delta_{l}}
\left(1+\frac{2 \Delta_l}{c} (z-w)^2 T(w)+ {\cal
O}\left((z-w)^3\right)\label{gcppope} \right),
\end{equation}
where $c=\frac{2(k-1)}{k+2}$ is the central charge of the
parafermion theory, $T$ the corresponding energy--momentum tensor
and the structure constants $C_{l_1,l_2}$, which are determined by
associativity of the OPE, are given by
\begin{equation}
C_{l_1,l_2}=\Bigg(\frac{\Gamma(k-l_1+1)\Gamma(k-l_2+1)\Gamma(l_1+l_2+1)}
{\Gamma(l_1+1)\Gamma(l_2+1)\Gamma(k+1)\Gamma(k-l_1-l_2+1)}\Bigg)^{\frac{1}{2}}.
\end{equation}
Using (\ref{gcpope}) and  (\ref{gcppope}) we find that
\begin{equation}
\label{cpOPE}
\begin{array}{rcl}
\displaystyle{\psi(z)\psi(w)
}&\sim&\displaystyle{(z-w)^{-\frac{2}{k}} \psi_2(w)} \ , \crbig
\displaystyle{\psi(z)\psi^\dagger(w)}&\sim&\displaystyle{
(z-w)^{-2\left(1-\frac{1}{k}\right)}}
\end{array}
\end{equation}
and upon using also
\begin{equation}
\label{POPE}
\begin{array}{rcl}
\displaystyle{\partial P(z) \partial P(w)
}&\sim&\displaystyle{-\frac{1}{(z-w)^2}} \ , \crbig
\displaystyle{e^{i a P(z)} e^{i b
P(w)}}&\sim&\displaystyle{(z-w)^{ab} e^{i a P(z)+i b P(w)}} \ ,
\end{array}
\end{equation}
we can check that the decomposed currents in (\ref{su2c2}) obey
the $SU(2)$ current algebra.

\no An affine $SU(2)$ primary $\Phi^{{ su}}_{j;m,\bar m}$
satisfies
\begin{equation}
\label{SUpri}
\begin{array}{rcl}
\displaystyle{J^3(z) \Phi^{{ su}}_{j;m,\bar
m}(w)}&\sim&\displaystyle{\frac{m}{z-w}\Phi^{{ su}}_{j;m,\bar m}}
\ , \crbig \displaystyle{J^\pm(z) \Phi^{{ su}}_{j;m,\bar
m}(w)}&\sim&\displaystyle{\frac{j\mp m}{z-w}\Phi^{{
su}}_{j;m\pm1,\bar m}}
\end{array}
\end{equation}
and decomposing it in terms of the parafermionic primary $\psi_{j;m,\bar m}$
as
\begin{equation}\label{SUpridecomp}
\Phi^{{ su}}_{j;m,\bar m}=\psi_{j;m,\bar m }\exp{i \left(m \sqrt{\frac{2}{k}} P_L +
\bar m \sqrt{\frac{2}{k}} P_R\right)}\ ,
\end{equation}
leads to the following OPEs for the primaries $\psi_{j;m,\bar m}$:
\begin{equation}
\label{cparapriOPE}
\begin{array}{rcl}
\displaystyle{\psi(z) \psi_{j;m,\bar m}(w)}&\sim&\displaystyle{
\frac{j-m}{\sqrt{k}} \frac{\psi_{j;m+1,\bar
m}}{(z-w)^{1+\frac{2m}{k}}}} \ , \crbig
\displaystyle{\psi^\dagger(z) \psi_{j;m,\bar
m}(w)}&\sim&\displaystyle{\frac{j+m}{\sqrt{k}}
\frac{\psi_{j;m-1,\bar m}}{(z-w)^{1-\frac{2m}{k}}}}\ .
\end{array}
\end{equation}

\subsection{Non-compact parafermions and $SL(2,\mathbb{R})$ current algebra}

We start now from the $SL(2,\mathbb{R})$ current algebra at level $k\geq 2$
\begin{equation}
\label{SLalg}
\begin{array}{rcl}
\displaystyle{K^3(z)
K^3(w)}&\sim&\displaystyle{-\frac{k}{2}\frac{1}{(z-w)^2}} \ ,
\crbig \displaystyle{K^3(z)K^\pm(w)}&\sim&\displaystyle{\pm
\frac{K^\pm(w)}{(z-w)}} \ , \crbig
\displaystyle{K^+(z)K^-(w)}&\sim&\displaystyle{\frac{k}{(z-w)^2}-
\frac{2 K^3(w)}{z-w}}
\end{array}
\end{equation}
and decompose the currents as
\begin{equation}
\label{SLcurdecomp}
\begin{array}{rcl}
\displaystyle{K^3}&=&\displaystyle{-\sqrt{\frac{k}{2}} \partial P}
\ , \crbig \displaystyle{K^{+}}&=&\displaystyle{\sqrt{k} \pi
\exp{\sqrt{\frac{2}{k}}P} } \ , \crbig
\displaystyle{K^{-}}&=&\displaystyle{\sqrt{k} \pi^\dagger \exp
{-\sqrt{\frac{2}{k}}P}  } \ ,
\end{array}
\end{equation}
with $P$ is a boson and $\pi,\pi^\dagger$ are the fundamental
non-compact parafermion fields.

\no The non-compact parafermionic algebra at level $k$ contains an
infinite set of objects $\pi_l$ and $\pi_l^\dagger=\pi_{-l}$ where
$l=0,1,2,\ldots$ and so that $\pi_0=1$ and $\pi_1=\pi$. Their
conformal dimensions are $\Delta_{l}=l+\frac{l^2}{k}$ and their
OPEs are the same as those of the compact parafermions but with
the central charge being $c=\frac{2(k+1)}{k-2}$ and the structure
constants $C_{l_1,l_2}$ changed to
\begin{equation}
C_{l_1,l_2}=\Bigg(\frac{\Gamma(k)\Gamma(k+l_1+l_2)\Gamma(l_1+l_2+1)}
{\Gamma(l_1+1)\Gamma(l_2+1)\Gamma(k+l_1)\Gamma(k+l_2)}\Bigg)^{\frac{1}{2}}.
\end{equation}
The OPEs we will need are
\begin{equation}
\label{ncpOPE}
\begin{array}{rcl}
\displaystyle{\pi(z)\pi(w)}&\sim&\displaystyle{
(z-w)^{\frac{2}{k}} \pi_2(w)} \ , \crbig
\displaystyle{\pi(z)\pi^\dagger(w)}&\sim&\displaystyle{
(z-w)^{-2(1+\frac{1}{k})}}\ .
\end{array}
\end{equation}

\no The $SL(2,\mathbb{R})$ affine primaries $\Phi^{sl}_{j;m, \bar m}$
satisfy
\begin{equation}
\label{}
\begin{array}{rcl}
\displaystyle{K^3(z) \Phi^{{ sl}}_{j;m, \bar m
}(w)}&=&\displaystyle{ \frac{m}{z-w} \Phi^{sl}_{j;m, \bar m}} \
, \crbig \displaystyle{K^\pm(z) \Phi^{{ sl}}_{j;m, \bar
m}(w)}&=&\displaystyle{ \frac{m\pm(j+1)}{z-w}
\Phi^{{sl}}_{j;m\pm1, \bar m}}
\end{array}
\end{equation}
and decomposing them as
\begin{equation}\label{SLpridecomp}
\Phi^{{ sl}}_{j;m, \bar m}=\pi_{j;m,\bar m}\exp \left({ m
\sqrt{\frac{2}{k}} P_{\mathrm{L}}+ \bar m  \sqrt{\frac{2}{k}}
P_{\mathrm{R}}}\right)\ ,
\end{equation}
leads to the following OPEs for the parafermionic primaries
$\pi_{j;m,\bar m}$:
\begin{equation}
\label{ncparapriOPE}
\begin{array}{rcl}
\displaystyle{\pi(z) \pi_{j;m,\bar m}(w)
}&\sim&\displaystyle{\frac{m+(j+1)}{\sqrt{k}}
\frac{\pi_{j;m+1,\bar m}}{(z-w)^{1-\frac{2m}{k}}}} \ , \crbig
\displaystyle{\pi^\dagger(z) \pi_{j;m,\bar
m}(w)}&\sim&\displaystyle{\frac{m-(j+1)}{\sqrt{k}}
\frac{\pi_{j;m-1,\bar m}}{(z-w)^{1+\frac{2m}{k}}}} \ .
\end{array}
\end{equation}

\section{\boldmath Superconformal ${\cal N}=2$ algebras
\unboldmath}\label{sucal}

Using the decomposition of the  ${\cal N}=2$ minimal model in
terms of the bosonic parafermion theory and a free scalar $P$, we
can write an explicit realization of the ${\cal N}=2$
superconformal algebra generators. They read:
\begin{equation}
\label{SUGpm}
\begin{array}{rcl}
\displaystyle{G^{ +su}}&=
&\displaystyle{\sqrt{\frac{2(k-2)}{k}}\psi^\dagger \exp {-i
\sqrt{\frac{k}{k-2}}P_{\mathrm{L}}}} \ , \crbig \displaystyle{G^{
-su}}&=&\displaystyle{\sqrt{\frac{2(k-2)}{k}} \psi \exp {+i
 \sqrt{\frac{k}{k-2}}P_{\mathrm{L}}}} \ ,
\end{array}
\end{equation}
 while the $R$-symmetry $U(1)$ current is
 \begin{equation}\label{SUJR}
 J^{ su}=-i \sqrt{\frac{k-2}{k}}\partial P\ .
 \end{equation}
Notice that when
the supersymmetric minimal model is at level $k$ the bosonic
parafermions are at level $k-2$.

\no
Similarly, for the  ${\cal N}=2$ Kazama--Suzuki model of $SL(2,\mathbb{R})/U(1)$ at level $k$
 the superconformal generators can be written in terms of the non-compact
 parafermions at level $k+2$ and of the free scalar $Q$ as
  \begin{equation}
\label{SLGpm}
\begin{array}{rcl}
\displaystyle{G^{ +sl}}&= &\displaystyle{\sqrt{\frac{2(k+2)}{k}}
\pi^\dagger \exp {i \sqrt{\frac{k}{k+2}}Q_{\mathrm{L}}}} \ ,
\crbig \displaystyle{G^{ -sl}}&= &\displaystyle{
\sqrt{\frac{2(k+2)}{k}} \pi \exp {-i
\sqrt{\frac{k}{k+2}}Q_{\mathrm{L}}}} \ ,
\end{array}
\end{equation}
while the $R$-symmetry $U(1)$ current is
 \begin{equation}\label{SUJR}
 J^{ sl}=i \sqrt{\frac{k+2}{k}}\partial Q\ .
 \end{equation}
It is straightforward to verify that these generators satisfy the
${\cal N}=2$ superconformal algebra by using the OPEs of the
parafermion theory provided in App. \ref{pOPE}.

\no
The superconformal generators of the total ${\cal N}=2$ algebra on
$SU(2)/U(1) \times SL(2,\mathbb{R})/U(1)$ are
\begin{equation}
G^+=G^{ +su}+G^{ +sl}, \;\;\;G^-=G^{ -su}+G^{ -sl}
\end{equation}
and the total $U(1)$ $R$-current is
\begin{equation}
J=J^{ su}+J^{ sl}.
\end{equation}
Analogous expressions hold for the antiholomorphic sector.

\section{\boldmath Coframes, spin connections and curvature two-forms \unboldmath}\label{geodata}

The full metric corresponding to small deformations of the circle is
given in \eqn{js79gen} which for reference we copy here
\begin{eqnarray}\label{D1}
 {ds^{2}\ov k}  &=&
 d\rho^2+\coth^2\rho \, d\om^2
+d\th^2 + \tan^2\th \, d\varphi^2
+2 \epsilon \frac{\sin^{n-2}\theta}{\cosh^{n}\r}
\times
\nonumber \\
&& \times\big[ {\cos n(\omega-\varphi)} \left(d\th^2-\tan^2\th \,
d\varphi^2\right)+2\sin n(\omega-\varphi) \tan\th\, d\varphi \,
d\th \big]\ .
\end{eqnarray}
The coframe we select is inspired by the form of the classical
parafermions in the unperturbed case and it reads:
\begin{equation}
\label{d.2}
\begin{array}{rcl}
\displaystyle{e^{\hat{1}}}&=&\displaystyle{\sqrt{k} \left(d\rho -
i \coth \rho\, d\omega \right)e^{-i\omega}} \ , \crbig
\displaystyle{e^{\hat{2}}}&=&\displaystyle{\sqrt{k} \left(d\rho +
i \coth \rho\, d\omega \right)e^{i\omega}} \ , \crbig
\displaystyle{e^{\hat{3}}}&=&\displaystyle{\sqrt{k}\left(d\th - i
\tan \theta\, d\varphi \right)e^{-i\varphi} +\epsilon
\frac{e^{-ni(\omega-\varphi)}\sin^{n-2}\theta}{\cosh^{n}\rho}
\sqrt{k} \left(d\th + i \tan \theta\, d\varphi
\right)e^{-i\varphi}} \ , \crbig
\displaystyle{e^{\hat{4}}}&=&\displaystyle{\sqrt{k}\left(d\th + i
\tan \theta\, d\varphi \right)e^{i\varphi} + \epsilon
\frac{e^{ni(\omega-\varphi)}\sin^{n-2}\theta}{\cosh^{n}\rho}
 \sqrt{k}\left(d\th - i \tan \theta\, d\varphi
\right)e^{i\varphi}} \ .
\end{array}
\end{equation}
We also define for convenience the unperturbed vielbeins in the compact directions
\begin{equation}
\label{vielbeins}
\begin{array}{rcl}
\displaystyle{e^{\hat{3}}_0}&=&\displaystyle{\sqrt{k}\left(d\th -
i \tan \theta\ d\varphi \right)e^{-i\varphi}} \ , \crbig
\displaystyle{e^{\hat{4}}_0 }&=&\displaystyle{\sqrt{k}\left(d\th +
i \tan \theta\, d\varphi \right)e^{i\varphi}} \ .
\end{array}
\end{equation}
The connection one-form
$
    \omega^{\hat \imath}_{\hphantom{\hat \imath}\hat \jmath} =
   \Gamma^{\hat \imath}_{\hphantom{\hat \imath}\hat{k}\hat \jmath}\,
   e^{\hat{k}}_{\vphantom{\hat \jmath}}$
is defined as usual by
\begin{equation}\label{dcoframe}
  de^{\hat \imath}_{\vphantom{\hat \jmath}} +
  \omega^{\hat \imath}_{\hphantom{\hat \imath}\hat \jmath}
  \wedge  e^{\hat \jmath}_{\vphantom{\hat \imath}}=0
\end{equation}
and the curvature two-form
\begin{equation}\label{curvform}
R^{\hat \imath}_{\hphantom{\hat \imath}\hat \jmath}= d\omega^{\hat
\imath}_{\hphantom{\hat \imath}\hat \jmath}+
 \omega^{\hat \imath}_{\hphantom{\hat \imath}\hat{k}} \wedge
\omega^{\hat{k}}_{\hphantom{\hat \imath}\hat \jmath}= \frac{1}{2}
R^{\hat \imath}_{\hphantom{\hat \imath}\hat
\jmath\hat{k}\hat{l}}\, e^{\hat{k}}_{\vphantom{\hat \jmath}}
\wedge e^{\hat{l}}_{\vphantom{\hat \imath}}\ .
\end{equation}
The non-vanishing components of the connection and curvature forms
corresponding to (\ref{d.2}) read
\begin{equation}
\label{connection}
\begin{array}{rcl}
\displaystyle{
\sqrt{k}\omega^{\hat{1}}_{\hphantom{\hat{\imath}}\hat{1}} =-
\sqrt{k}\omega^{\hat{2}}_{\hphantom{\hat{\imath}}\hat{2}}}&=&\displaystyle{-\frac{\coth
\rho}{2}\left[e^{i\omega} e^{\hat{1}}_{\vphantom{\hat{\jmath}}}
-e^{-i\omega} e^{\hat{2}}_{\vphantom{\hat{\jmath}}} \right]} \ ,
\crbig
\displaystyle{\sqrt{k}\omega^{\hat{3}}_{\hphantom{\hat{\imath}}\hat{3}}
=- \sqrt{k}\omega^{\hat{4}}_{\hphantom{\hat{\imath}}\hat{4}}
}&=&\displaystyle{ \frac{\tan \th}{2}\left[e^{i\varphi}
e^{\hat{3}}_{\vphantom{\hat{\jmath}}} -e^{-i\varphi}
e^{\hat{4}}_{\vphantom{\hat{\jmath}}} \right]} \crbig
&&\displaystyle{+ \frac{\epsilon}{2} \frac{\tan \th \,
\sin^{n-2}\theta}{\cosh^{n}\r}\left[-
e^{ni(\omega-\varphi)+i\varphi} \,
e^{\hat{3}}_{\vphantom{\hat{\jmath}}}+
e^{-ni(\omega-\varphi)-i\varphi}  \,
e^{\hat{4}}_{\vphantom{\hat{\jmath}}} \right]} \ , \crbig
\displaystyle{\sqrt{k}\omega^{\hat{1}}_{\hphantom{\hat{\imath}}\hat{4}}
=- \sqrt{k}\omega^{\hat{3}}_{\hphantom{\hat{\imath}}\hat{2}}}&=&
\displaystyle{\epsilon n   \frac{\tanh\rho \sin^{n-2}\theta
}{\cosh^{n}\r}e^{-ni(\omega-\varphi)-2i\varphi-i \omega}\,
e^{\hat{4}}_{\vphantom{\hat{\jmath}}}} \ , \crbig
\displaystyle{\sqrt{k}\omega^{\hat{2}}_{\hphantom{\hat{\imath}}\hat{3}}
=- \sqrt{k}\omega^{\hat{4}}_{\hphantom{\hat{\imath}}\hat{1}}
}&=&\displaystyle{ \epsilon n \frac{\tanh\rho\sin^{n-2}\theta
}{\cosh^{n}\r} e^{ni(\omega-\varphi)+2i\varphi+i \omega} \,
e^{\hat{3}}_{\vphantom{\hat{\jmath}}}}
\end{array}
\end{equation}
and
\begin{equation}
\label{curv2form}
\begin{array}{rcl}
\displaystyle{ kR^{\hat{1}}_{\hphantom{\hat{\imath}}\hat{1}}
=-kR^{\hat{2}}_{\hphantom{\hat{\imath}}\hat{2}}}&=&\displaystyle{-\frac{1}{\sinh^2\rho}\,
e^{\hat{1}}_{\vphantom{\hat{\jmath}}}\wedge
e^{\hat{2}}_{\vphantom{\hat{\jmath}}}} \ ,  \crbig
\displaystyle{kR^{\hat{3}}_{\hphantom{\hat{\imath}}\hat{3}}
=-kR^{\hat{4}}_{\hphantom{\hat{\imath}}\hat{4}}
}&=&\displaystyle{-\frac{1}{\cos^2\th}\,
e^{\hat{3}}_{\vphantom{\hat{\jmath}}}\wedge
e^{\hat{4}}_{\vphantom{\hat{\jmath}}}}
   \crbig
&&\displaystyle{+ \epsilon\ n \frac{\tan\theta \sin^{n-2}\theta
\tanh\rho}{\cosh^{n} \rho} e^{ni(\omega-\varphi)+i\varphi+i\omega
}\, e^{\hat{1}}_{\vphantom{\hat{\jmath}}}\wedge
e^{\hat{3}}_{\vphantom{\hat{\jmath}}} } \crbig &&\displaystyle{-
\epsilon\ n \frac{\tan\theta \sin^{n-2}\theta \tanh\rho}{\cosh^{n}
\rho} e^{-ni(\omega-\varphi)-i\varphi-i\omega }\,
e^{\hat{2}}_{\vphantom{\hat{\jmath}}}\wedge
e^{\hat{4}}_{\vphantom{\hat{\jmath}}}}
  \crbig
&&\displaystyle{+ 2 \epsilon \frac{\tan^2\theta
\sin^{n-2}\theta}{\cosh^{n} \rho} \cos n(\omega-\varphi)\,
e^{\hat{3}}_{\vphantom{\hat{\jmath}}}\wedge
e^{\hat{4}}_{\vphantom{\hat{\jmath}}}} \ ,  \crbig
\displaystyle{kR^{\hat{1}}_{\hphantom{\hat{\imath}}\hat{4}}=-kR^{\hat{3}}_{\hphantom{\hat{\imath}}\hat{2}}}
&=&\displaystyle{\epsilon\ n \frac{\left(1 -n \sinh^2
\rho\right)\sin^{n-2}\theta}{\cosh^{n+2} \rho}
e^{-ni(\omega-\varphi)-2i\varphi-2i\omega }\,
e^{\hat{2}}_{\vphantom{\hat{\jmath}}}\wedge
e^{\hat{4}}_{\vphantom{\hat{\jmath}}}}
  \crbig
&&\displaystyle{ + \epsilon \ n \frac{\tan\theta \sin^{n-2}\theta
\tanh\rho}{\cosh^{n} \rho}e^{-ni(\omega-\varphi)-i\varphi-i\omega
}\, e^{\hat{3}}_{\vphantom{\hat{\jmath}}}\wedge
e^{\hat{4}}_{\vphantom{\hat{\jmath}}}} \ ,  \crbig
\displaystyle{kR^{\hat{2}}_{\hphantom{\hat{\imath}}\hat{3}}
=-kR^{\hat{4}}_{\hphantom{\hat{\imath}}\hat{1}}}&=&\displaystyle{\epsilon\
n \frac{\left(1-n \sinh^2 \rho\right)\sin^{n-2}\theta}{\cosh^{n+2}
\rho} e^{ni(\omega-\varphi)+2i\varphi+2i\omega }\,
e^{\hat{1}}_{\vphantom{\hat{\jmath}}}\wedge
e^{\hat{3}}_{\vphantom{\hat{\jmath}}}}
  \crbig
&&\displaystyle{- \epsilon\ n \frac{\tan\theta \sin^{n-2}\theta
\tanh\rho}{\cosh^{n} \rho}e^{ni(\omega-\varphi)+i\varphi+i\omega
}\, e^{\hat{3}}_{\vphantom{\hat{\jmath}}}\wedge
e^{\hat{4}}_{\vphantom{\hat{\jmath}}}} \ .
\end{array}
\end{equation}

\end{document}